\documentclass[useAMS,usenatbib]{mnras}
\usepackage{graphicx}
\usepackage{amsmath}
\usepackage{amssymb}
\usepackage{color}

\def \etal {et~al.~}

\def \spose#1{\hbox  to 0pt{#1\hss}}  
\def \lta{\mathrel{\spose{\lower 3pt\hbox{$\sim$}}\raise  2.0pt\hbox{$<$}}}
\def \gta{\mathrel{\spose{\lower  3pt\hbox{$\sim$}}\raise 2.0pt\hbox{$>$}}}

\newcommand{\Mpch}{{\ifmmode{h^{-1}{\rm Mpc}}\else{$h^{-1}$Mpc}\fi}}
\newcommand{\Msun}{{\ifmmode{ {\rm M}_{\odot}}\else{M_{\odot}}\fi}}
\newcommand{\pc}{{\ifmmode{ {\rm pc}^{-2}}\else{pc$^{-2}$}\fi}}
\newcommand{\kms}{{\ifmmode{ {\rm km\,s^{-1}} }\else{ ${\rm km\,s^{-1}}$ }\fi}}
\def \hi {{\sc Hi} }
\newcommand{\Vmax}{\ifmmode{V_{\rm max}}\else{$V_{\rm max}$}\fi}
\newcommand{\VmaxDMO}{\ifmmode{V_{\rm max}^{\rm DMO}}\else{$V_{\rm max}^{\rm DMO}$}\fi}
\newcommand{\VHI}{\ifmmode{V_{\rm HI}}\else{$V_{\rm HI}$}\fi}
\newcommand{\fbar}{\ifmmode{f_{\rm bar}}\else{$f_{\rm bar}$}\fi}

\def \Omegam {\ifmmode \Omega_{\rm m} \else $\Omega_{\rm m}$ \fi} 
\def \Omegab {\ifmmode \Omega_{\rm b} \else $\Omega_{\rm b}$ \fi} 
\def \fbar {\ifmmode f_{\rm bar} \else $f_{\rm bar}$ \fi} 
\def \OmegaL {\ifmmode \Omega_{\rm \Lambda} \else $\Omega_{\rm \Lambda}$\fi} 
\def \Deltavir {\ifmmode \Delta_{\rm vir} \else $\Delta_{\rm vir}$ \fi}
\def \rhocrit {\ifmmode \rho_{\rm crit} \else $\rho_{\rm crit}$ \fi}

\title[The diversity of dwarf galaxy kinematics]{NIHAO XVII: The diversity of dwarf galaxy kinematics and implications for the \hi velocity function}
\author[Dutton \etal]{Aaron A. Dutton$^1$\thanks{dutton@nyu.edu},
  Aura Obreja$^{1,2}$,   Andrea V. Macci\`{o}$^{1,3}$\\
 $^{1}$New York University Abu Dhabi, PO Box 129188, Saadiyat Island, Abu Dhabi, United Arab Emirates\\
  $^{2}$University Observatory Munich, Scheinerstra\ss e 1, D-81679 Munich, Germany\\
  $^{3}$Max Planck Institute f\"{u}r Astronomie, K\"{o}nigstuhl 17, 69117 Heidelberg, Germany\\
}
\begin{document}

\maketitle

\label{firstpage}

\begin{abstract}
We use 85 pairs of high resolution LCDM cosmological simulations from
the NIHAO project to investigate why in dwarf galaxies neutral
hydrogen (HI) linewidths measured at 50\% of the peak flux $W_{50}/2$
(from the hydrodynamical simulations) tend to underpredict the maximum
circular velocity $\VmaxDMO$ (from the corresponding dark matter only
simulations).  There are two main contributing processes.  1) Lower
mass galaxies are less rotationally supported. This is confirmed
observationally from the skewness of linewidths in bins of HI mass in
both ALFALFA and HIPASS observations.  2) The HI distributions are
less extended (relative to the dark matter halo) in dwarf galaxies.
Coupled to the lower baryon-to-halo ratio results in rotation curves
that are still rising at the last measured data point, in agreement
with observations from SPARC.  Combining these two effects, in both
simulations and observations lower mass galaxies have, on average,
lower $W_{50}/W_{20}$.  Additionally,  mass loss driven by supernovae
and projection effects (dwarf galaxies are in general not thin disks)
further reduce the linewidths.   The implied HI linewidth velocity
function from NIHAO is in good agreement with observations in the
nearby Universe of dwarf galaxies: $10 < W_{50}/2 < 80 \,\kms$. The
dark matter only slope of $\approx -2.9$ is reduced to $\approx -1.0$
in the hydro simulations.  Future radio observations of unbiased
samples with higher spatial resolution will enable stricter tests of
the simulations, and thus of the LCDM model.
\end{abstract}


\vspace{-0.4cm}
\section{Introduction}

The Lambda Cold Dark Matter (LCDM) model is very successful in
reproducing the large scale structure of the Universe
\citep{Springel05} and the anisotropies in the Cosmic Microwave
Background \citep{Planck14}.  In a LCDM universe there is a larger
number of low mass structures compared to more massive ones,  or in
other words a steeply declining halo mass function $N(M) \propto
M^{-1}$ or halo velocity function $N(V)\propto V^{-3}$
\citep{Klypin15}.  This prediction is naively at odds with
observational data around galaxies, i.e. the satellite abundance
\citep{Klypin99, Moore99}.  This ``missing satellite'' problem has
natural baryonic solution: galaxy formation becomes increasingly
inefficient in low-mass dark matter haloes, due to the ionizing
background, supernovae (SN) explosions and gas removal due to ram
pressure \citep{Bullock00,Benson02,Kravtsov04,Maccio10}.  The latest
cosmological hydrodynamical simulations of the Local Group are now
consistent with the observed satellite stellar mass and velocity
functions \citep[e.g.,][]{Sawala16,Buck18}.

The velocity function of galaxies using HI linewidths has been
measured in the nearby universe (within $\sim 200$ Mpc) from the
HIPASS \citep{Zwaan10} and ALFALFA surveys \citep{Papastergis11}, and
within the Local Volume (within $10$ Mpc) by \citet{Klypin15}. These
and other authors \citep{Zavala09, Trujillo-Gomez11}  have shown that
LCDM provides very good estimates of the number of galaxies with
circular velocities around and above $80\,\kms$.  With a careful
treatment of various systematics including survey selection effects
\citet{Obreschkow13} showed agreement between LCDM down to circular
velocities of $\sim 60 \kms$.  However, LCDM appears to fail  quite
dramatically at lower circular velocities, overestimating by a factor
of $\simeq 3$ the number of dwarf galaxies at velocity scale of
$50\,\kms$, and by a factor of $\simeq 5$ at velocity scale of
$30\,\kms$ \citep{Klypin15}.   Galaxies at these velocity scales are
essentially insensitive to the ionization background and, by not being
satellites, they are not affected by gas depletion via ram pressure or
by stellar stripping. This makes the mismatch between the observed
linewidth function and the halo velocity function a more serious
challenge to the LCDM paradigm. Furthermore, simple modifications to
the LCDM paradigm, such as warm dark matter, are unable to resolve
this issue as candidates with a warm enough particle to significantly
reduce the number density of dark matter haloes result in a
characteristic scale in the velocity function that is not observed
\citep{Klypin15}. The normalization of the velocity function is
dependent on key cosmological parameters such as the amplitude of
primordial perturbations ($\sigma_8$) and the total matter density
($\Omegam$) \citep{Schneider18}. For example, the difference between a
Planck \citep{Planck14} and WMAP7 \citep{WMAP7} cosmology is a factor
of 1.43  in number density at fixed velocity \citep{Klypin15}.

The solution to this discrepancy is that in dwarf galaxies HI
linewidths measured at 50\% of the peak flux, $W_{50}$, are a poor
tracer of the maximum circular velocity of the host dark matter halo
\citep{Brook15, Maccio16, Brooks17, Verbeke17}.  This should not be a
surprise, since it is generally known that lower mass and younger
galaxies are less rotationally supported than higher mass, and older
galaxies. This is seen in both observations \citep[e.g.][]{Kassin12,
  Newman13, Wisnioski15}, and hydrodynamical galaxy formation
simulations \citep{Ceverino17, El-Badry18a}.  Furthermore, as
discussed in \citet{Courteau97}, unlike resolved \hi rotation curves,
\hi linewidths do not necessarily sample outer disks effectively since
the \hi surface density drops rapidly beyond the optical radius.

In this paper we investigate the cause of the difference between
$W_{50}/2$ and $\VmaxDMO$ further using galaxy formation simulations
from the NIHAO project \citep{Wang15}.  Possible explanations we
consider include: gas mass loss from the halo -- resulting in lower
circular velocity in the hydro than dark matter only simulation; HI in
the simulations is not extended enough -- so that the maximum rotation
velocity does not reach the maximum circular velocity; Non-circular
motions -- resulting in the rotation curve underestimating the
circular velocity; Projection effects -- due to dwarf galaxies not
being thin rotating disks.

The outline of the paper is as follows. Section \ref{sec:sims}
describes the cosmological hydrodynamical simulations and the various
velocity definitions we use. In Section \ref{sec:vmaxw} we present the
transformation from maximum circular velocity to HI linewidth. In
section \ref{sec:vf} we present the implied HI velocity function from
our simulations. In section \ref{sec:tests} we discuss what causes the
linewidths to underestimate the maximum halo velocity, and present
observational tests of these effects. A summary in given in section
\ref{sec:sum}.

\begin{figure*}
  \includegraphics[width=0.45\textwidth]{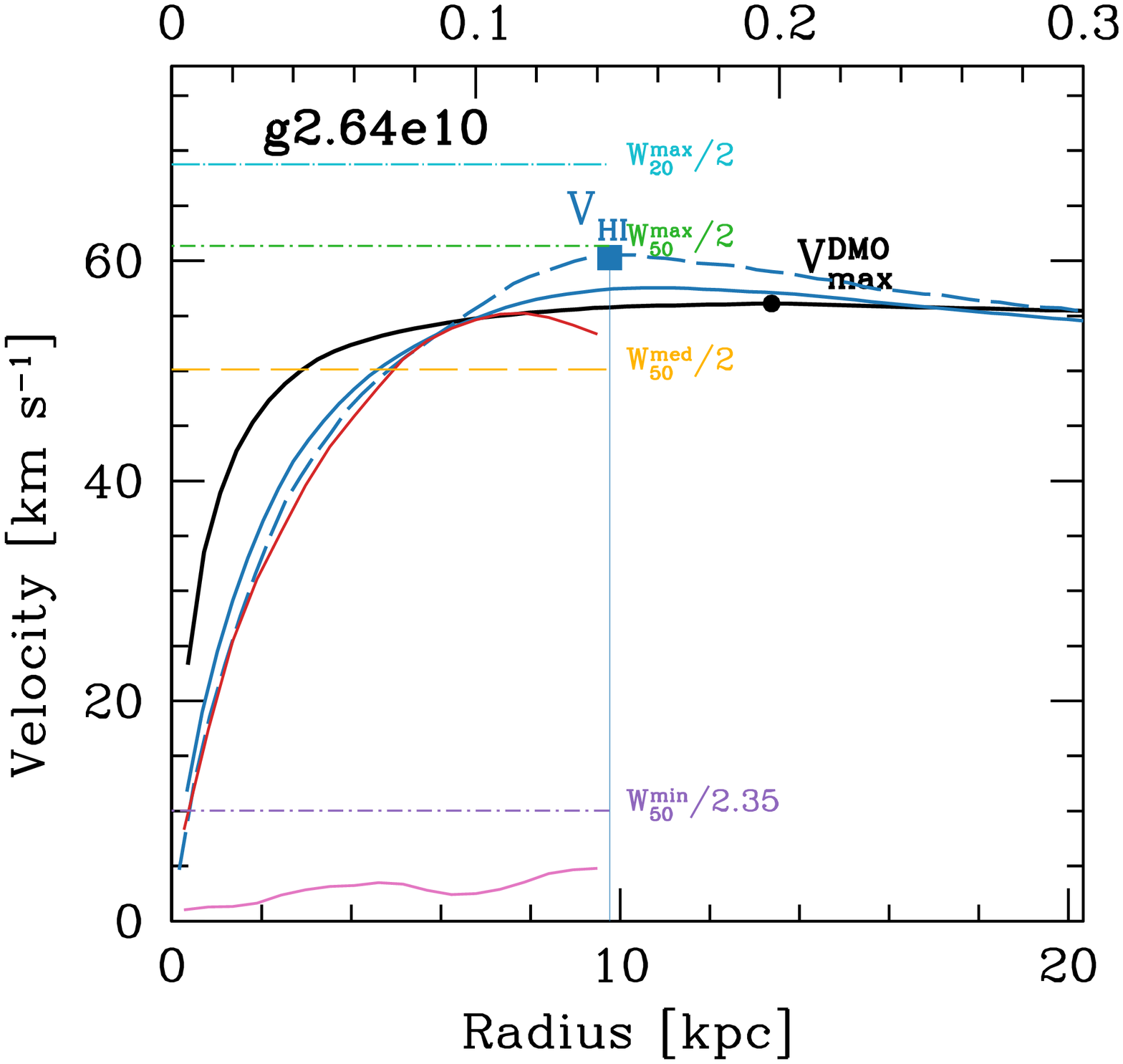}
  \includegraphics[width=0.45\textwidth]{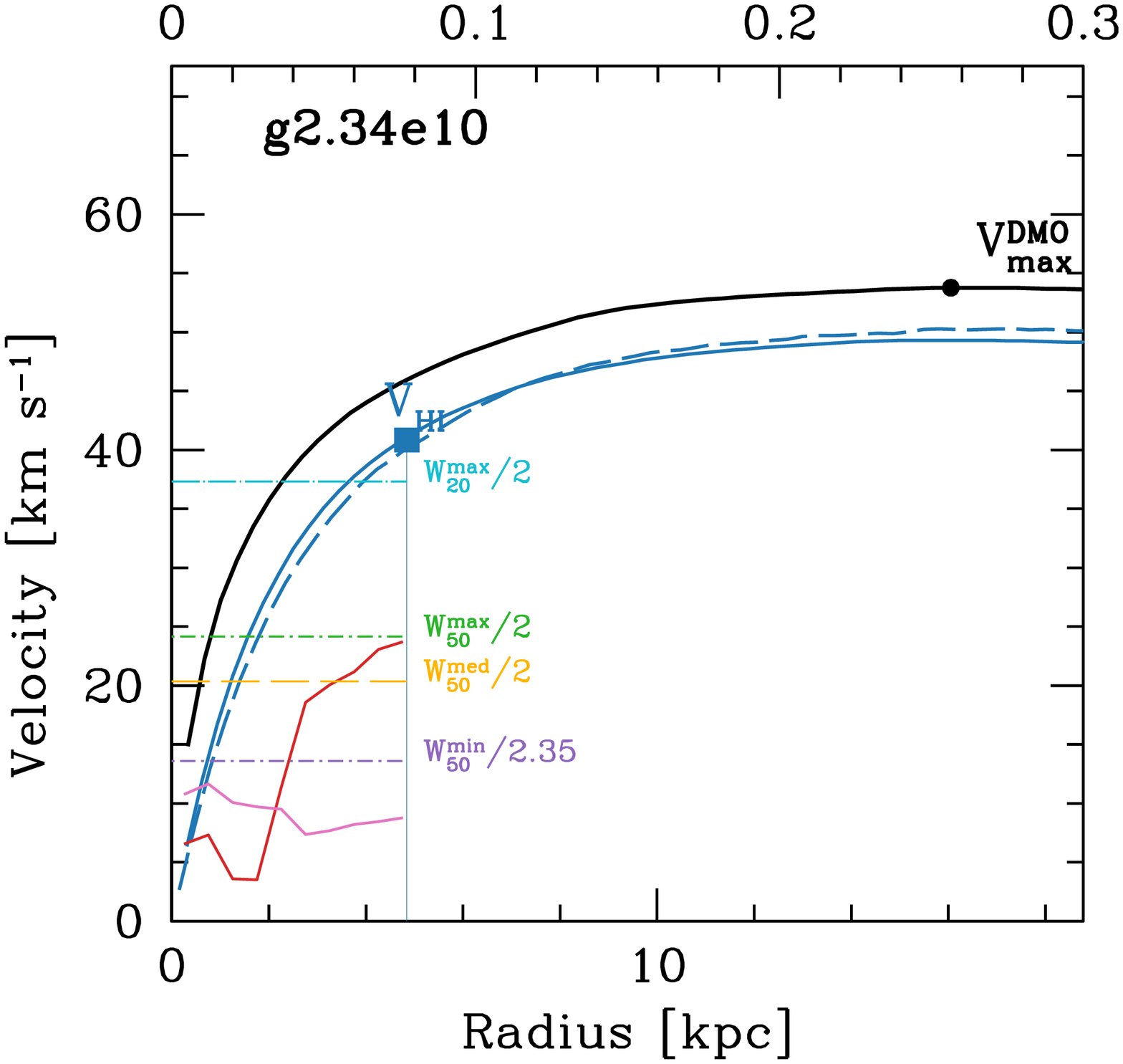}
  \includegraphics[width=0.45\textwidth]{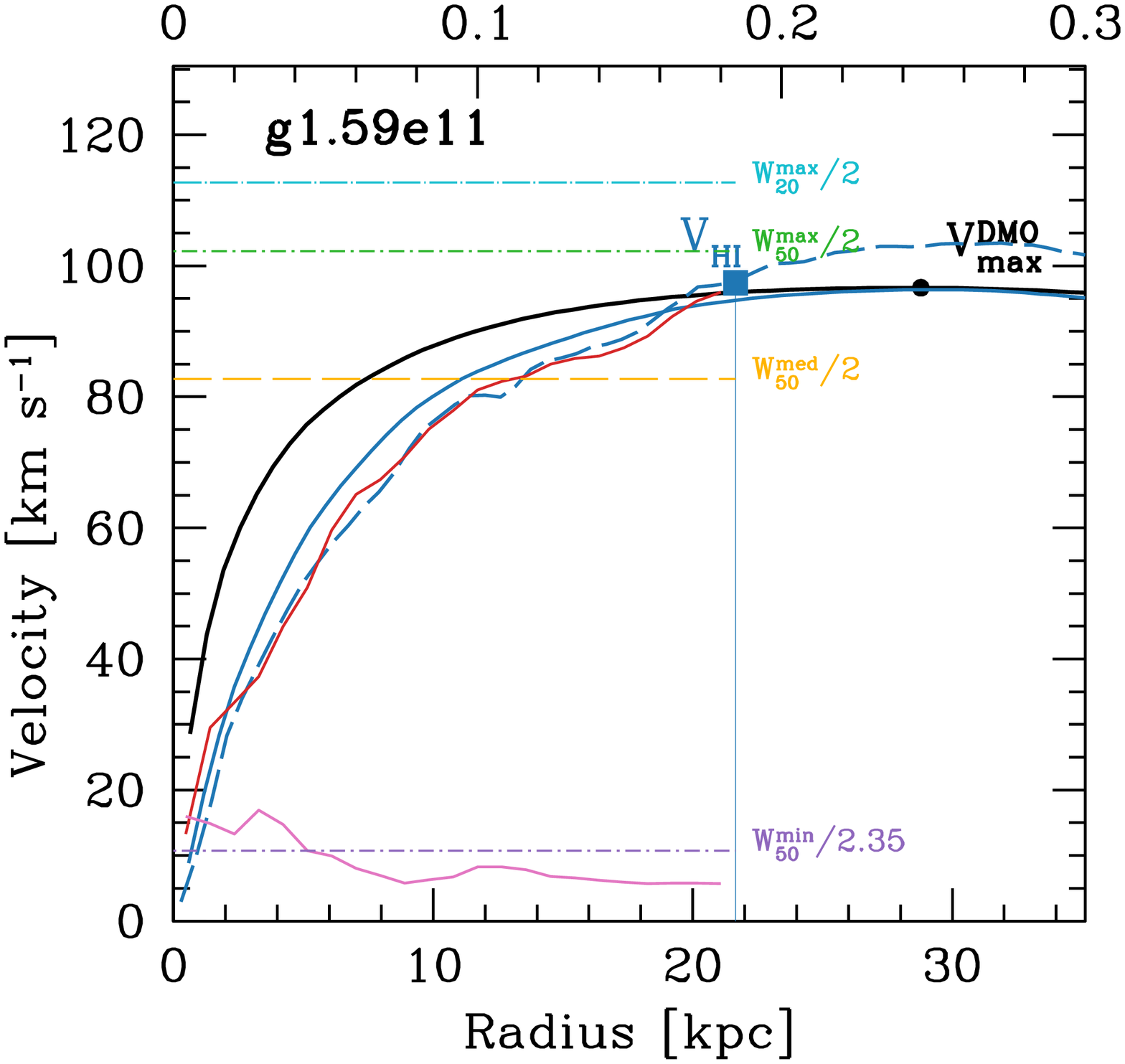}
  \includegraphics[width=0.45\textwidth]{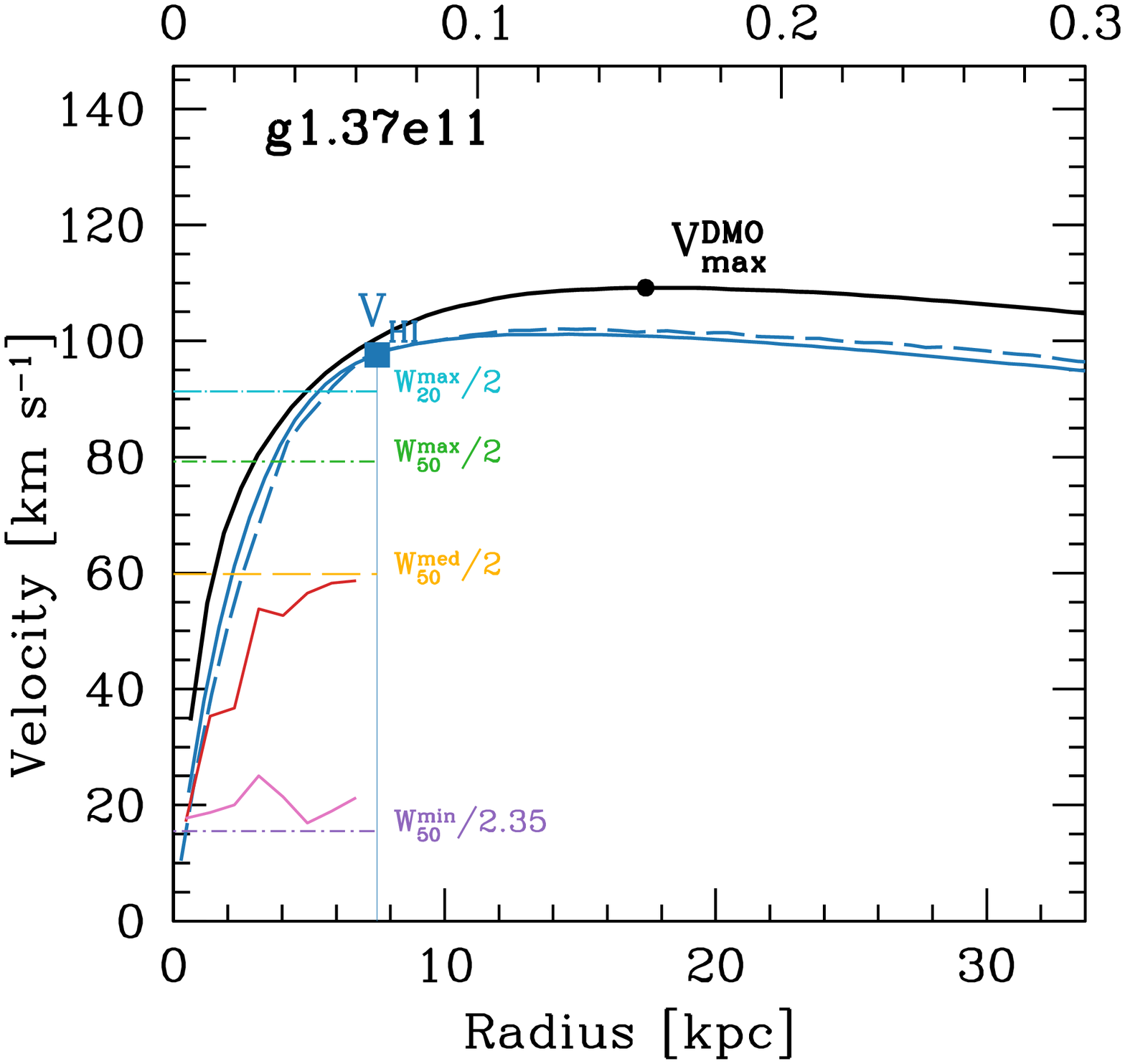}
\caption{Example velocity profiles of four galaxies illustrating the
  different velocity definitions.  The black line shows the spherical
  circular velocity of the dark matter only simulation. The black
  circle shows the maximum of this curve, $\VmaxDMO$, which typically
  occurs at $\approx 20\%$ of the virial radius, $R_{200}$, (upper
  axis).  The two galaxies in the upper panels have $\VmaxDMO\sim
  55\kms$, while the galaxies in the lower panels have $\VmaxDMO\sim
  100 \kms$. The blue lines show the circular velocity of the
  hydrodynamical simulation using the spherically enclosed mass (solid
  lines) and the potential in the disk plane (dashed).  The left
  panels show galaxies where the rotation curve (red) traces the
  circular velocity from the potential (blue-dashed), and the maximum
  linewidth $W_{50}/2$ (green dashed line) equals the circular
  velocity at the HI radius, $V_{\rm HI}$ (blue square). The right
  panels show galaxies where the rotation curve underpredicts the
  circular velocity, and $W_{50}/2$ underestimates $V_{\rm HI}$. The
  pink lines show the bulk velocity dispersion of the \hi gas. A
  global measurement of the gas dispersion (including the thermal
  broadening) is given by the minimum HI linewidth / 2.35 (purple
  lines).} 
\label{fig:vr}
\end{figure*}

\section{Simulations}
\label{sec:sims}

Here we give a brief overview of the NIHAO simulations. We refer the
reader to \citet{Wang15} for a more complete discussion.  NIHAO  is a
sample of $\sim$ 90 hydrodynamical cosmological zoom-in simulations
using the SPH code {\sc gasoline2} \citep{Wadsley17}.

Haloes are selected at redshift $z=0$ from parent dissipationless
simulations of size 60, 20, \& 15 Mpc$/h$, presented in
\citet{Dutton14} which adopt a flat $\Lambda$CDM cosmology with
parameters from the \citet{Planck14}: Hubble parameter $H_0$= 67.1
\kms Mpc$^{-1}$, matter density $\Omegam=0.3175$, dark energy density
$\Omega_{\Lambda}=1-\Omegam=0.6825$, baryon density $\Omegab=0.0490$,
power spectrum normalization $\sigma_8 = 0.8344$, power spectrum slope
$n=0.9624$.  Haloes are selected uniformly in log halo mass from $\sim
10$ to $\sim 12$ {\it without} reference to the halo merger history,
concentration or spin parameter.  Star formation and feedback is
implemented as described in \citet{Stinson06, Stinson13}.  Mass and
force softening are chosen to resolve the mass profile at $\lta 1\%$
the virial radius, which results in $\sim 10^6$ dark matter particles
inside the virial radius of all haloes at $z=0$. The motivation of
this choice is to ensure that the simulations resolve the galaxy
dynamics on the scale of the half-light radii, which are typically
$\sim 1.5\%$ of the virial radius \citep{Kravtsov13}. Hydro particles
have force softenings a factor of 2.34 smaller than the dark matter
particles, and range from $\simeq 75 {\rm pc}$ in the lowest mass
haloes to $\simeq 400 {\rm pc}$ in the most massive haloes.

Each hydro simulation has a corresponding dark matter only (DMO)
simulation of the same resolution. These simulations have been started
using the identical initial conditions, replacing baryonic particles
with dark matter particles.  We remove the four  most massive haloes
(g1.12e12, g1.77e12, g1.92e12, g2.79e12), as these have formed too
many stars, in particular near the galaxy centers, resulting in
strongly peaked central circular velocity profiles \citep{Wang15}.
The final sample used in this work consists of 85 simulations.

Haloes in NIHAO zoom-in simulations were identified using the
MPI+OpenMP hybrid halo finder
\texttt{AHF}\footnote{http://popia.ft.uam.es/AMIGA} \citep{Gill04,
  Knollmann09}. \texttt{AHF} locates local over-densities in an
adaptively smoothed density field as prospective halo centers. The
virial masses of the haloes are defined as the masses within a sphere
whose average density is 200 times the cosmic critical matter density,
$\rhocrit=3H_0^2/8\pi G$.  The virial mass, size and circular velocity
of the hydro simulations are denoted: $M_{200}, R_{200}, V_{200}$.
The corresponding properties for the dark matter only simulations are
denoted with a superscript, ${\rm DMO}$.  For the baryons we calculate
masses enclosed within spheres of radius $r_{\rm gal}=0.2R_{200}$,
which corresponds to $\sim 10$ to $\sim 50$ kpc.  The stellar mass
inside $r_{\rm gal}$ is $M_{\rm star}$, the neutral hydrogen, inside
$r_{\rm gal}$ is computed following \citet{Rahmati13} as described in
\citet{Gutcke17}. In principle the neutral hydrogen should be
separated further into atomic (H{\sc i}), and molecular, ($H_2$),
gas. In practice this has only marginal impact on the derived \hi
linewidths. This is expected because in dwarf galaxies the ratio
between molecular and atomic gas is very low, while in high mass
galaxies both atomic and molecular gas trace the same flat rotation
velocity, albeit from gas at different radii.  As a fiducial choice we
consider atomic gas to be neutral gas with a density less than
$10\,{\rm cm^{-3}}$ (which is also the star formation threshold used
in our simulations).  We measure galaxy velocities using a number of
definitions as discussed below.

The NIHAO simulations are the largest set of cosmological zoom-ins
covering the halo mass range $10^{10}$ to $10^{12}\Msun$. Their
uniqueness is in the combination of high spatial resolution coupled to
a statistical sample of haloes. In the context of LCDM they form the
``right'' amount of stars both today and at earlier times
\citep{Wang15}. Their cold gas masses and sizes are consistent with
observations \citep{Stinson15, Maccio16}, they follow the gas,
stellar, and baryonic Tully-Fisher relations \citep{Dutton17}, and
they reproduce the diversity of dwarf galaxy rotation curve shapes
\citet{Santos-Santos18}.  As such they provide a good template with
which to connect galaxy observables (such as HI linewidths) to
intrinsic properties of the host dark matter halo (such as maximum
circular velocity).

\begin{figure*}
  \includegraphics[width=0.45\textwidth]{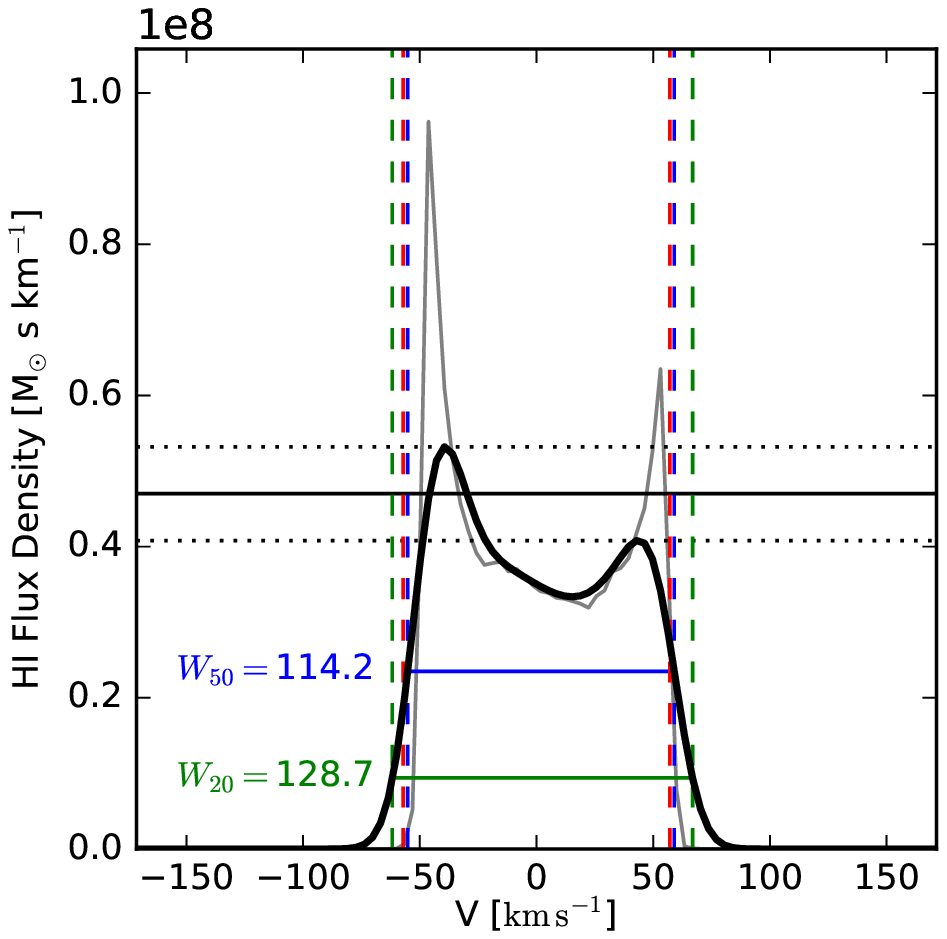}
  \includegraphics[width=0.45\textwidth]{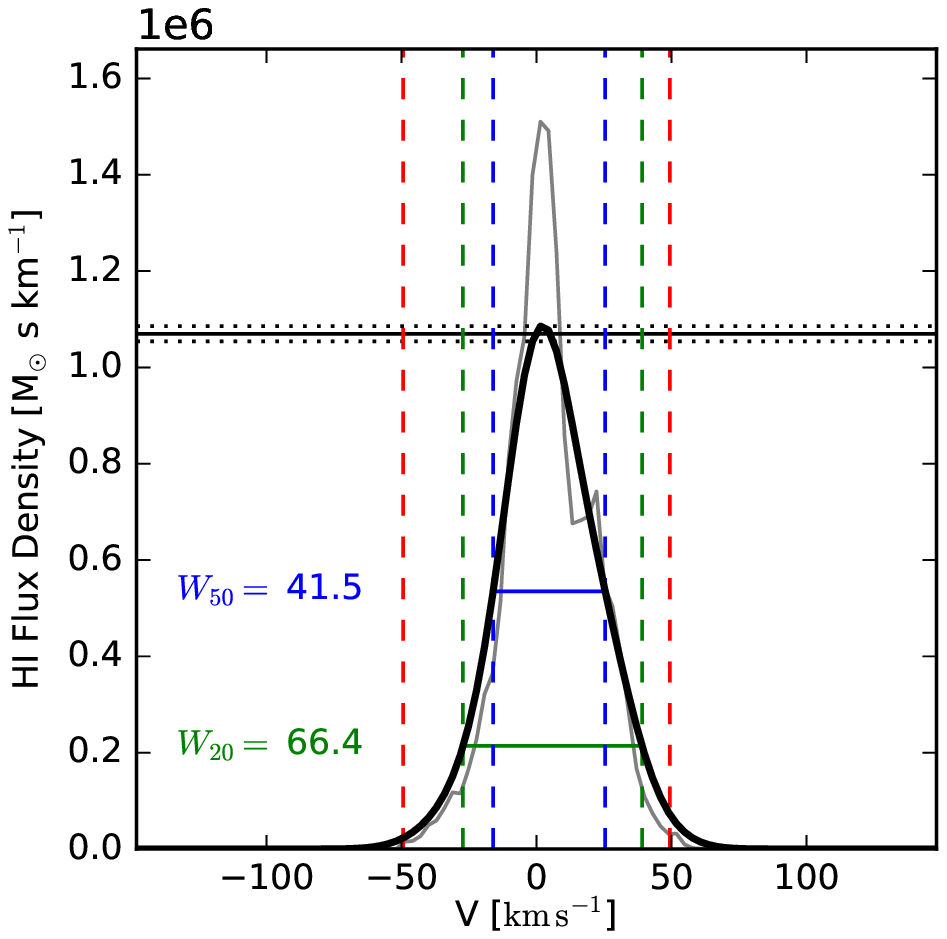}
  \includegraphics[width=0.45\textwidth]{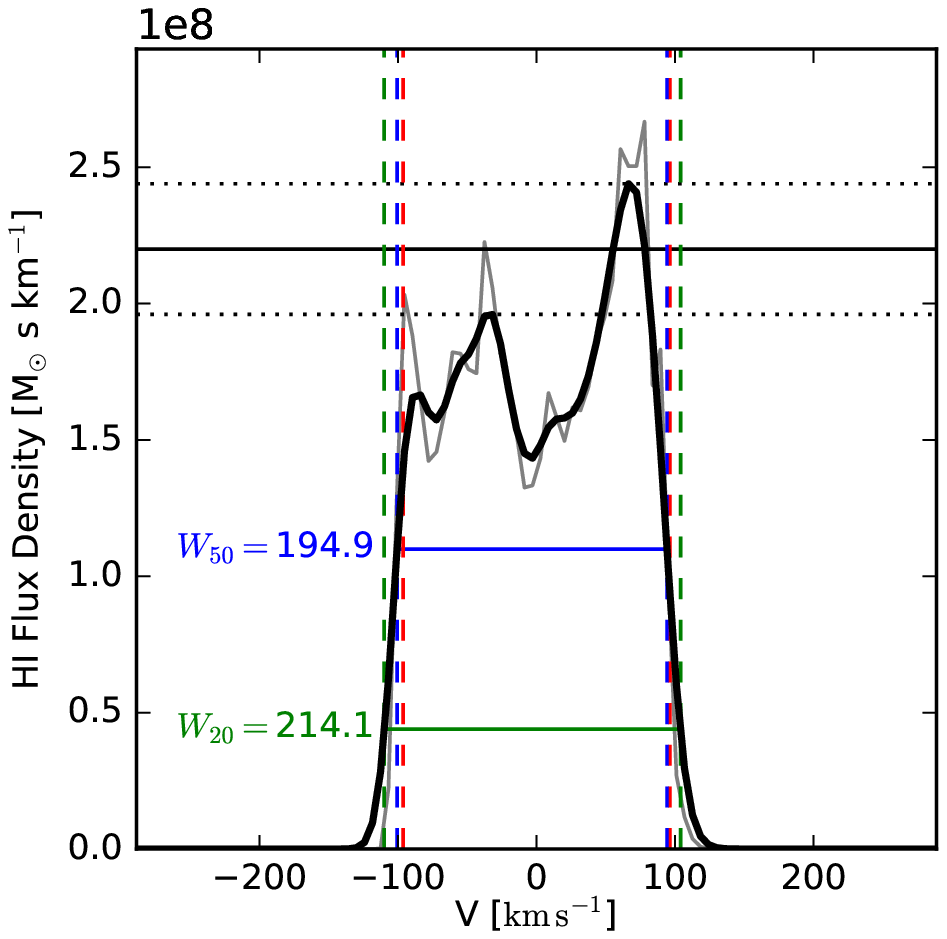}
  \includegraphics[width=0.45\textwidth]{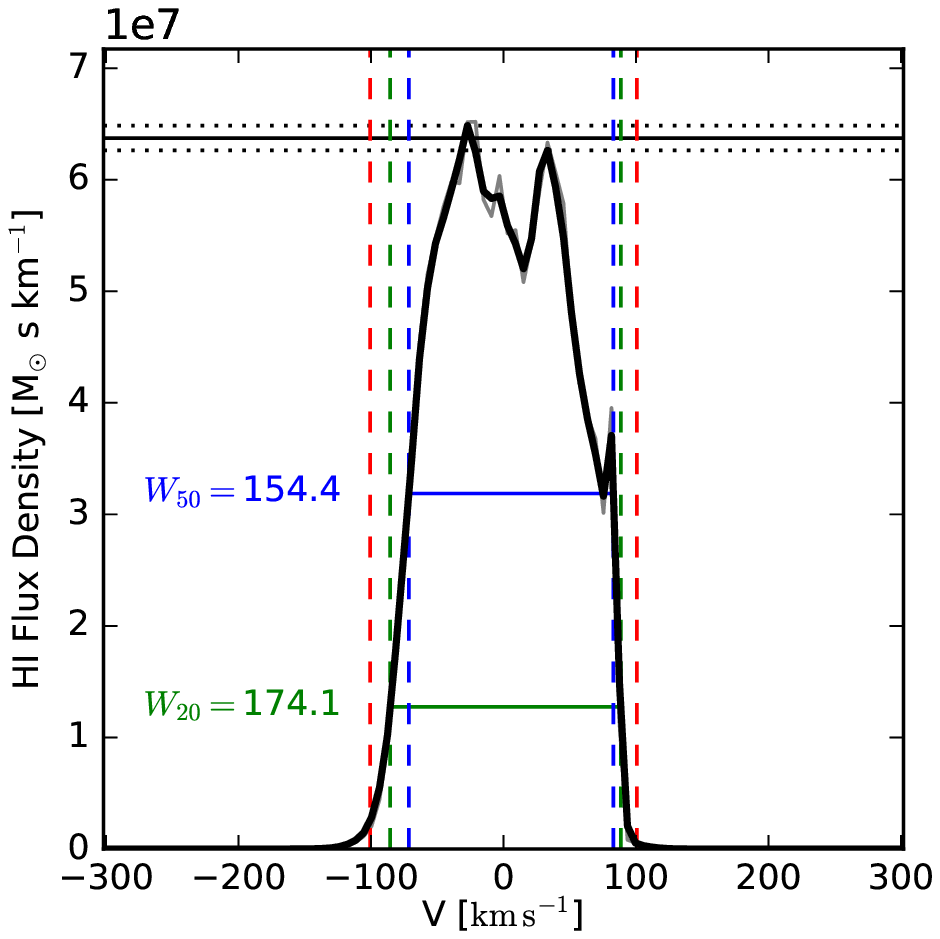}

\caption{Example HI line profiles for edge-on projections of the four
  galaxies from Fig.~\ref{fig:vr} illustrating the different linewidth
  definitions.  The black line shows the line profile including the
  thermal broadening, while the thin grey line shows the line profile
  from the kinematics only. The differences are only significant for
  the low velocity haloes. The horizontal dashed lines show the peak
  fluxes from either velocity side, while the horizontal solid line
  shows the average peak flux. The 50 and 20\% peak fluxes are given
  with blue and green horizontal lines respectively. The vertical blue
  and green dashed lines show the corresponding velocities at the 50
  and 20\% peak fluxes. The red vertical line shows the maximum
  circular velocity of the dark matter halo, $\VmaxDMO$.}
\label{fig:w}
\end{figure*}

\subsection{Velocity definitions}
We use a number of velocity definitions in this paper. We define them
with the help of some examples as shown in Fig.~\ref{fig:vr}.  These
galaxies were chosen to represent both rotationally supported galaxies
(left panels) and those with more pressure support (right panels).

\begin{itemize}
  \item {\bf $\VmaxDMO$ -- maximum circular velocity of the DMO simulation.}
We start with the spherical circular velocity profile of the dark
matter only simulation (solid black line), which is equivalent to the
cumulative mass profile  $M(<r) = r\, V(r)^2/G$.  The maximum spherical
circular velocity is shown as a black circle,
and occurs at 15\% to 25\% of the virial radius, $R_{200}$ (upper axis scale).

\item {\bf $V_{\rm max}$ -- maximum circular velocity of hydro simulation.}
We next consider the circular velocity profile of the hydrodynamical
simulation (blue lines). The solid line shows the spherical circular
velocity, while the dashed line shows that derived from the potential
in the disk plane, $V^2_{\rm pot}=-R {\partial\Phi/\partial R}$.
We define $\Vmax\equiv V_{\rm circ}(R_{\rm
  max}^{\rm DMO})$ as the spherical circular velocity of the hydro
simulation at the radius where the maximum circular velocity of the
DMO simulation occurs. We use this definition as it explicitly shows
when there is mass loss from the system ($\Vmax<\VmaxDMO$), and where
there is baryon dissipation ($\Vmax>\VmaxDMO$). In these examples,
there is mass loss from the galaxies in the right panels in Fig.~\ref{fig:vr},
and no change for the galaxies in the left panels.

Notice that the potential based circular velocity is lower than the
spherical circular velocity at small radii, and higher at large
radii. This is a well known feature of the differences between a
flattened and spherical mass distribution \citep{Binney87}

\begin{figure*}
  \includegraphics[width=0.48\textwidth]{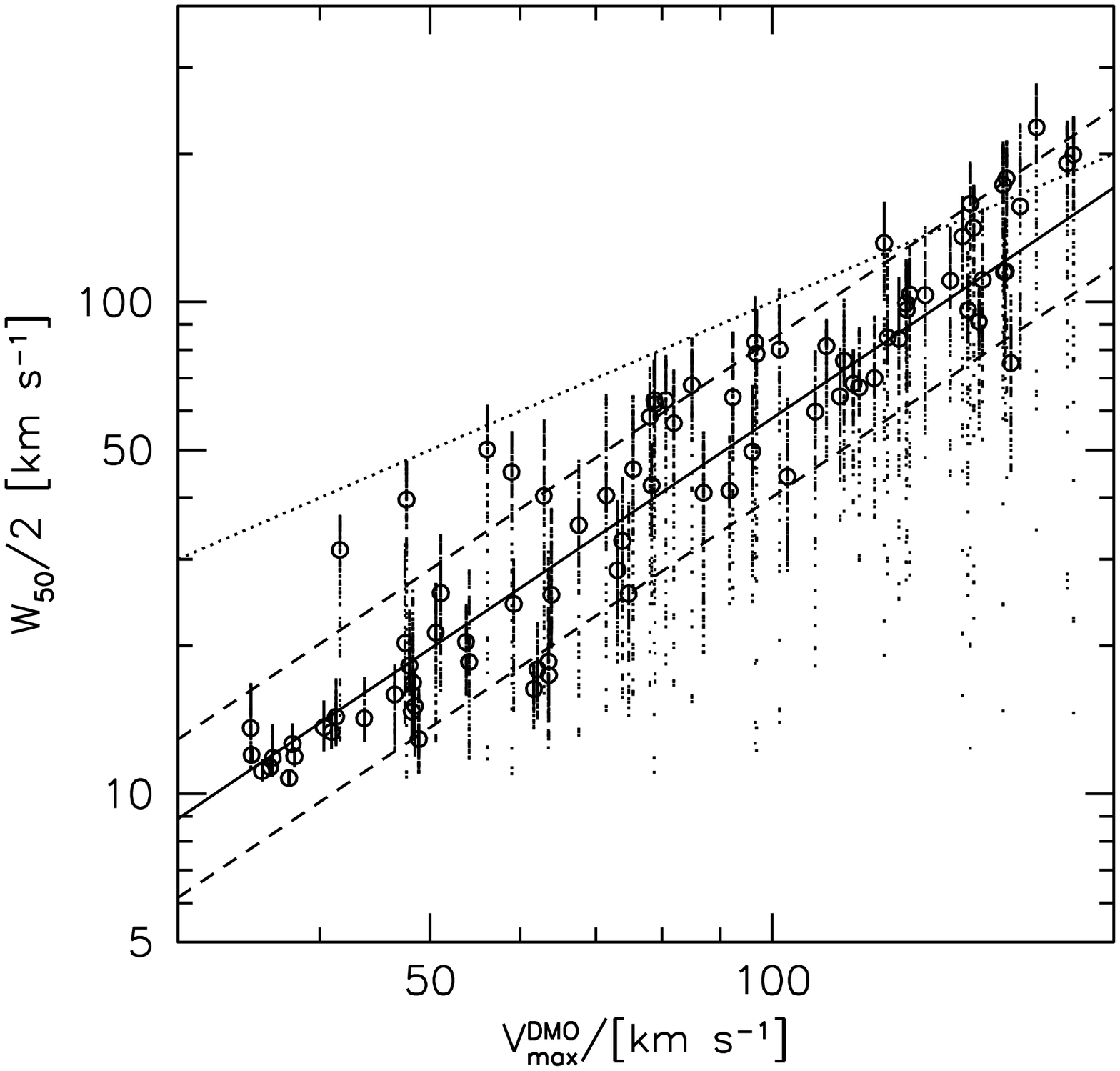}
  \includegraphics[width=0.48\textwidth]{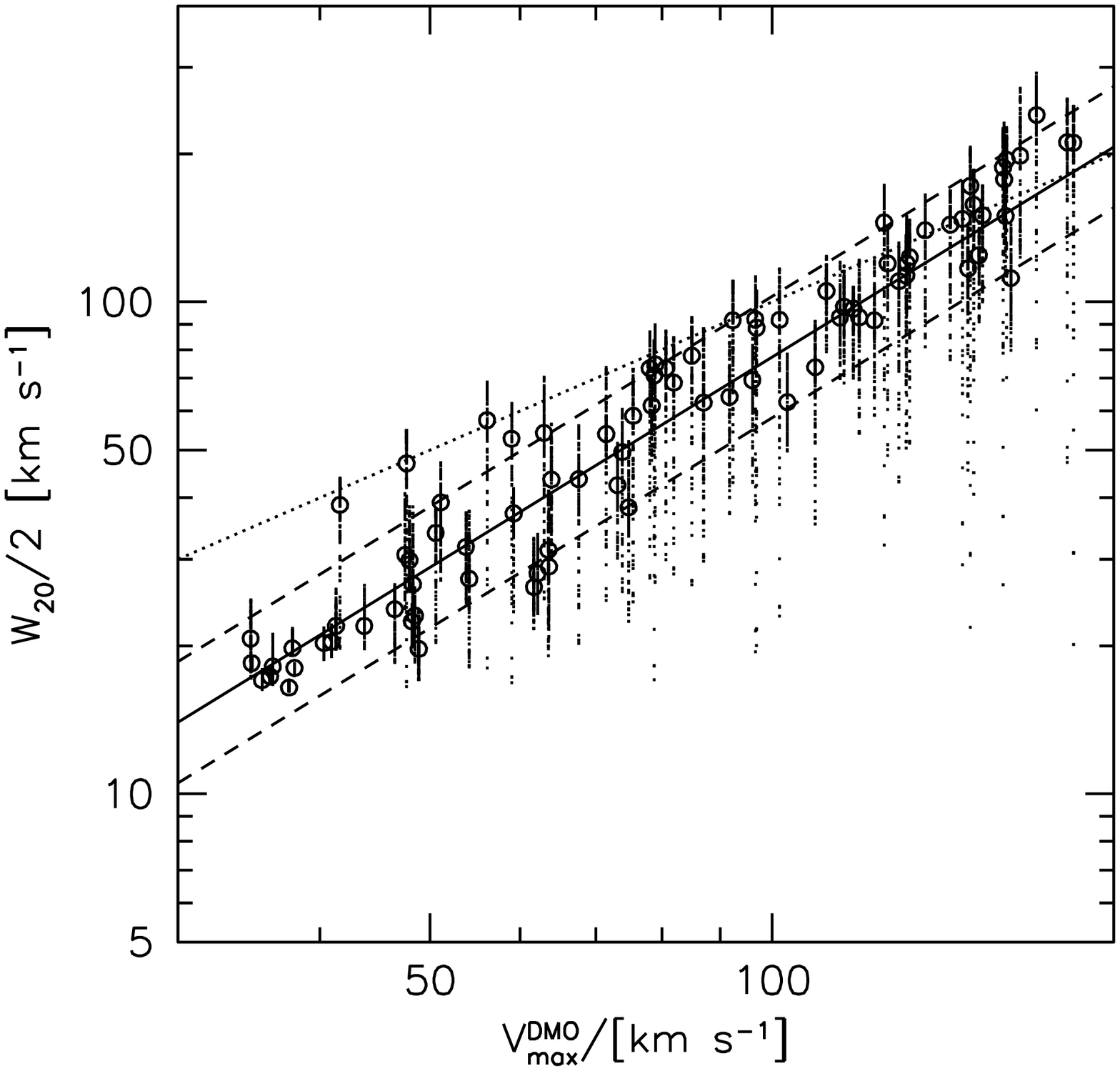}
  \caption{Relation between \hi linewidths, $W_{50}$ (left), and
  $W_{20}$ (right) from the hydro simulations and maximum circular
  velocity from the corresponding dark matter only simulations. The
  points show 100 random projections per galaxy, with the open circle
  showing the median linewidth.  The dotted line shows the one-to-one
  relation. The solid line shows a fit to all the data, with the
  $1\sigma$ scatter shown with dashed lines.}
\label{fig:wvmax}
\end{figure*}

\item {\bf $V_{\rm HI}$ - circular velocity at HI radius.}  We define
  the HI radius, $R_{\rm HI}$, as enclosing 90\% of the HI mass (in
  the face-on projection). This approximates the edge of the
  observable HI rotation curve, and occurs close to the radius where
  the projected HI surface density profile reaches $1 \Msun {\rm
    pc}^{-2}$, but it has the advantage of being free from projection
  effects and is measurable in every galaxy (that contains HI).  The
  potential based circular velocity at the HI radius is then, $V_{\rm
    HI}\equiv V_{\rm circ}(R_{\rm HI})$.

\item {\bf $V_{\rm rot}(R), \sigma(R)$ -- rotation and dispersion
  profile.}  We calculate the rotation curve (red line), and
  dispersion profile (pink line) as follows. We rotate the galaxy to a
  face-on view, using the angular momentum of the cold ($T<15000 K$)
  gas particles inside 10\% of the virial radius. We then divide the
  galaxy into a series of annuli of equal width. In each ring we
  calculate the mean velocity ($V_{\rm rot}$) and velocity dispersion
  ($\sigma$) in the azimuthal ($\phi$) direction. Note that this
  dispersion only takes into account the bulk motions of the gas.  In
  the galaxies in the left panels, the rotation velocity closely traces the
  potential based circular velocity, and the dispersion is low, $\sim
  5-10 \kms$. In the galaxies in the right panels, the rotation velocity
  systematically underpredicts the circular velocity, and the velocity
  dispersion is higher, $\sim 10-20 \kms$.

\item $W_{50}, W_{20}$ - {\bf HI linewidths}.  We next consider the HI
  linewidths, measured at 50\% and 20\% of the peak flux (see
  Fig.~\ref{fig:w} for examples from our reference galaxies).  In an
  update to the calculation in \citet{Maccio16} here we include the
  thermal motions of the HI gas. At large linewidths the changes are
  minor, but for small linewidths the temperature adds a floor to the
  linewidth of $~\sim 20\, \kms$, corresponding to $\sigma \sim 8.5\,
  \kms$.  For each gas particle we represent the line-of-sight
  velocity distribution with a Gaussian with mean corresponding to the
  line-of-sight velocity of the particle, a flux corresponding to the
  HI mass of the particle, and a dispersion computed from the
  temperature of the particle assuming $\sigma=\sqrt{k_{\rm B}T/m_{\rm
      H}}=9.09 [\kms] \sqrt{T/10000 {\rm K}}$.  We sum up the
  Gaussians to give the flux vs velocity histogram.
  
  As is commonly done in observations, we define the peak flux as the
  average of the peak flux on each velocity wing (i.e., positive and
  negative). See Fig.~\ref{fig:w} for examples. The peak fluxes on
  each velocity wing are marked with dotted horizontal lines. The
  average of these is shown with solid horizontal lines.  By
  definition $W_{20} \ge W_{50}$.  Because $W_{20}$ requires higher
  signal to noise, and thus can be measured from fewer galaxies,
  $W_{50}$ is the more commonly used definition in large HI surveys,
  so will be our default definition.

  We calculate the linewidths from 100 random projections (uniformly
  selected from the surface of a unit sphere), and measure the
  maximum, median, and minimum values. The maximum values typically
  correspond to the values derived from an edge-on view of the
  galaxy. They are shown as cyan ($W_{20}^{\rm max}$) and green
  ($W_{50}^{\rm max}$) horizontal lines in Fig.~\ref{fig:vr}.  The
  median $W_{50}/2$ is shown with an orange horizontal line. The
  minimum linewidth (purple horizontal lines) can be used to
  approximate the gas velocity dispersion of the system, since for a
  Gaussian, $\sigma = W_{50}/2.35$. 

\end{itemize}

For each projection where we measure a linewidth we also measure the minor
to major axis ratio (b/a) of the HI gas using moments as follows.  For
each gas particle we have its projected coordinates, $(x, y)$,
distance from the galaxy center $R^2=x^2+y^2$, and HI mass, $m$. Using
these quantities we calculate the moments weighted by the HI mass:
\begin{eqnarray}
  Sxx=\left(\sum m x^2/R^2 \right) / \sum m,\\
  Syy=\left(\sum m y^2/R^2 \right) / \sum m,\\
  Sxy=\left(\sum m xy/R^2 \right) / \sum m.  
\end{eqnarray}
The axis ratio is then given by
\begin{equation}
(b/a)=\frac{1-\sqrt{Q^2 + U^2}}{1+\sqrt{Q^2 + U^2}},
\end{equation}
where $Q=Sxx-Syy$ and $U=2Sxy$.

\begin{figure*}
  \includegraphics[width=0.90\textwidth]{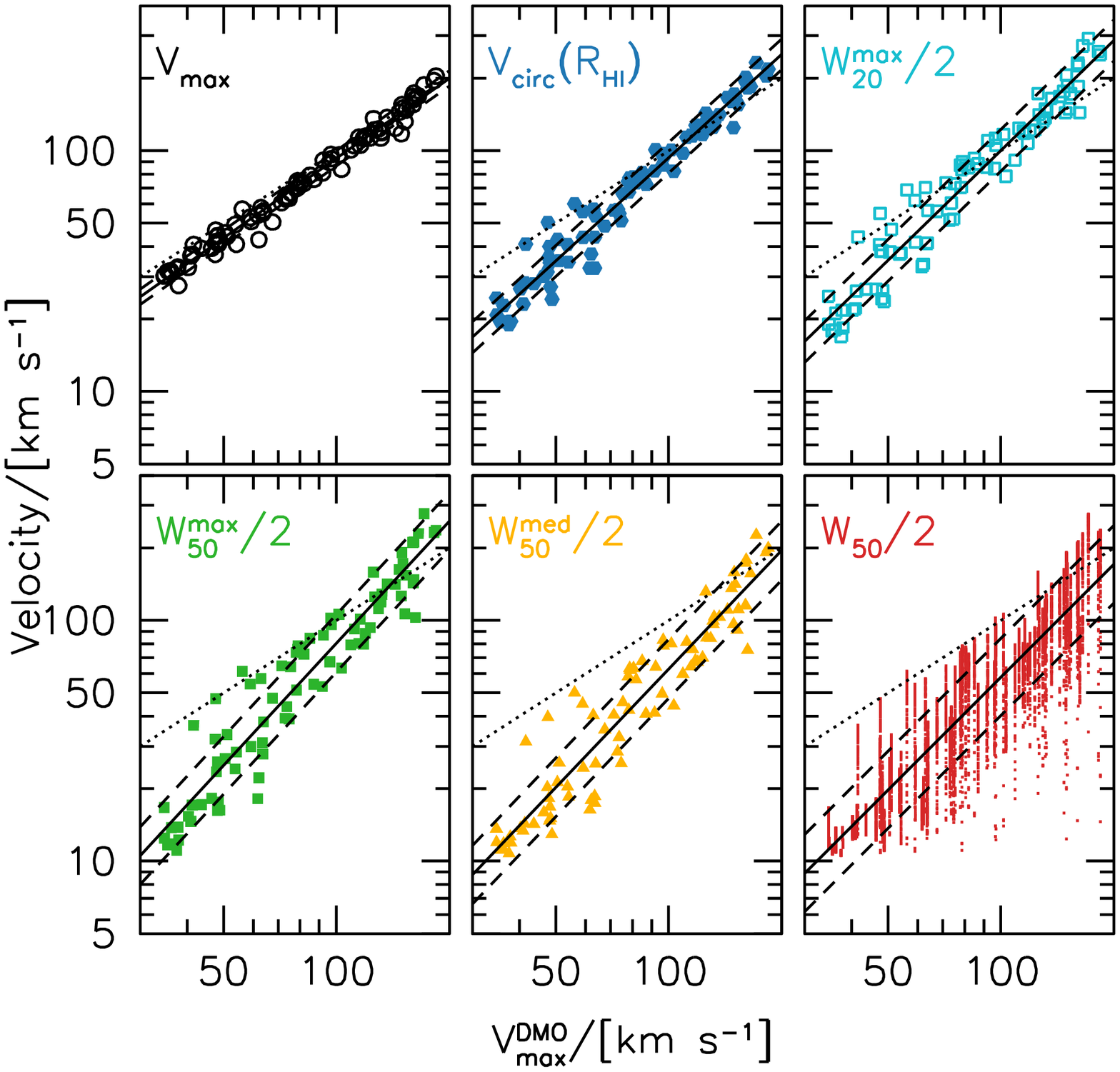}
\caption{From maximum halo velocity $\VmaxDMO$ to projected linewidth
  $W_{50}/2$ for the full sample. Each panel shows a velocity vs
  $\VmaxDMO$ relation. A fit to each relation of the form
  $y=a+b(x-x_0)$ is shown with solid (mean) and dashed (standard
  deviation) lines. The parameters of the fits are given in
  Table~\ref{tab:fits}. For reference, the one-to-one line is shown
  with a dotted line.  We see that the slope and scatter get
  progressively larger as we go from top left to bottom right. }
\label{fig:vvmax2}
\end{figure*}

\section{From maximum circular velocity to HI linewidth}
\label{sec:vmaxw}

Fig.~\ref{fig:wvmax} shows the relations between half-linewidths
$W_{50}/2$ and $W_{20}/2$ with $\VmaxDMO$.  The left panel is very
similar to Fig.2 of \citet{Maccio16}, but here we use an updated
calculation of the linewidth which introduces a floor to the linewidth
of $W_{50}\sim 20\kms$.  Galaxies lie systematically below the
one-to-one relation (dotted lines) indicating the \hi linewidth is a
biased tracer of the maximum halo velocity.  Linear fits to the
relations (using all the data points) are given by
\begin{equation}
  \log(W_{50}/2)= 1.614 +1.557(\log\VmaxDMO -1.904),
\end{equation}
with a scatter of $\sigma=0.16$, and
\begin{equation}
  \log(W_{20}/2)= 1.751 +1.419(\log\VmaxDMO -1.904),
\end{equation}
with a scatter of $\sigma=0.12$.  The slope of these fits is greater
than unity, showing that linewidths are a poorer tracer of halo
velocity in lower velocity haloes. The scatters show that $W_{50}$ is
a poorer tracer of the halo velocity than $W_{20}$.

\begin{table*}
\begin{center}
  \caption{Linear fits of the form $y=a+b(x-x_0)$ to various scaling
    relations. For all relations $x_0$ is chosen to equal the mean of
    $x$. The scatter of the data about the model is $\sigma_y$.} 
\begin{tabular}{cccccccc}
  \hline
  \hline
Sample & $y$ & $x$ & $x_0$ & $a$ & $b$ & $\sigma_{y}$ & Fig.\\
\hline 
NIHAO & $\log_{10}(W_{50}/2[\kms])$         & $\log_{10}(V_{\rm max}^{\rm DMO}/[\kms])$ & 1.904 & 1.614 & 1.557 & 0.161 & 3\\
NIHAO & $\log_{10}(W_{20}/2[\kms])$         & $\log_{10}(V_{\rm max}^{\rm DMO}/[\kms])$ & 1.904 & 1.751 & 1.419 & 0.120 & 3\\
NIHAO & $\log_{10}(V_{\rm max}/[\kms])$      & $\log_{10}(V_{\rm max}^{\rm DMO}/[\kms])$ & 1.904 & 1.863 & 1.101 & 0.032 & 4\\
NIHAO & $\log_{10}(V_{\rm HI}/[\kms])$       & $\log_{10}(V_{\rm max}^{\rm DMO}/[\kms])$ & 1.904 & 1.833 & 1.426 & 0.064 & 4\\
NIHAO & $\log_{10}(W_{20}^{\rm max}/2[\kms])$ & $\log_{10}(V_{\rm max}^{\rm DMO}/[\kms])$ & 1.904 & 1.855 & 1.518 & 0.086 & 4\\
NIHAO & $\log_{10}(W_{50}^{\rm max}/2[\kms])$ & $\log_{10}(V_{\rm max}^{\rm DMO}/[\kms])$ & 1.904 & 1.742 & 1.685 & 0.119 & 4\\
NIHAO & $\log_{10}(W_{50}^{\rm med}/2[\kms])$ & $\log_{10}(V_{\rm max}^{\rm DMO}/[\kms])$ & 1.904 & 1.640 & 1.635 & 0.122 & 4\\
\hline
SPARC & $\log_{10}(R_1/[\rm kpc])$        & $\log_{10}(M_{\rm star}/\Msun)$ & 9.493 & 1.109 & 0.281 & 0.18 & 15\\
NIHAO & $\log_{10}(R_1/[\rm kpc])$        & $\log_{10}(M_{\rm star}/\Msun)$ & 8.325 & 0.779 & 0.303 & 0.24 & 15\\
SPARC & $\log_{10}(R_{\rm last}/[\rm kpc])$ & $\log_{10}(M_{\rm star}/\Msun)$ & 9.471 & 1.109 & 0.311 & 0.23 & 15\\
NIHAO & $\log_{10}(R_{90}/[\rm kpc])$      & $\log_{10}(M_{\rm star}/\Msun)$ & 8.309 & 0.789 & 0.252 & 0.23 & 15\\
\hline
SPARC & $Vslope$                         & $\log_{10}(M_{\rm star}/\Msun)$ & 9.471 & 0.123 & -0.137 & 0.19 & 16\\
NIHAO & $Vslope$                         & $\log_{10}(M_{\rm star}/\Msun)$ & 8.394 & 0.286 & -0.098 & 0.14 & 16\\
SPARC & $Vslope$                     & $\log_{10}(R_{\rm last}/[\rm kpc])$ & 1.109 & 0.123 & -0.399 & 0.18 & 16\\
NIHAO & $Vslope$                     & $\log_{10}(R_{90}/[\rm kpc])$      & 0.826 & 0.286 & -0.332 & 0.15 & 16\\
\hline
NIHAO & $(b/a)_{\rm HI}$                   & $\log_{10}(M_{\rm HI}/\Msun)$   & 8.83 & 0.585 & -0.013 & 0.20 & 17\\
 \hline
 \hline
\end{tabular}
\label{tab:fits}
\end{center}
\end{table*}

\begin{figure*}
  \includegraphics[width=0.90\textwidth]{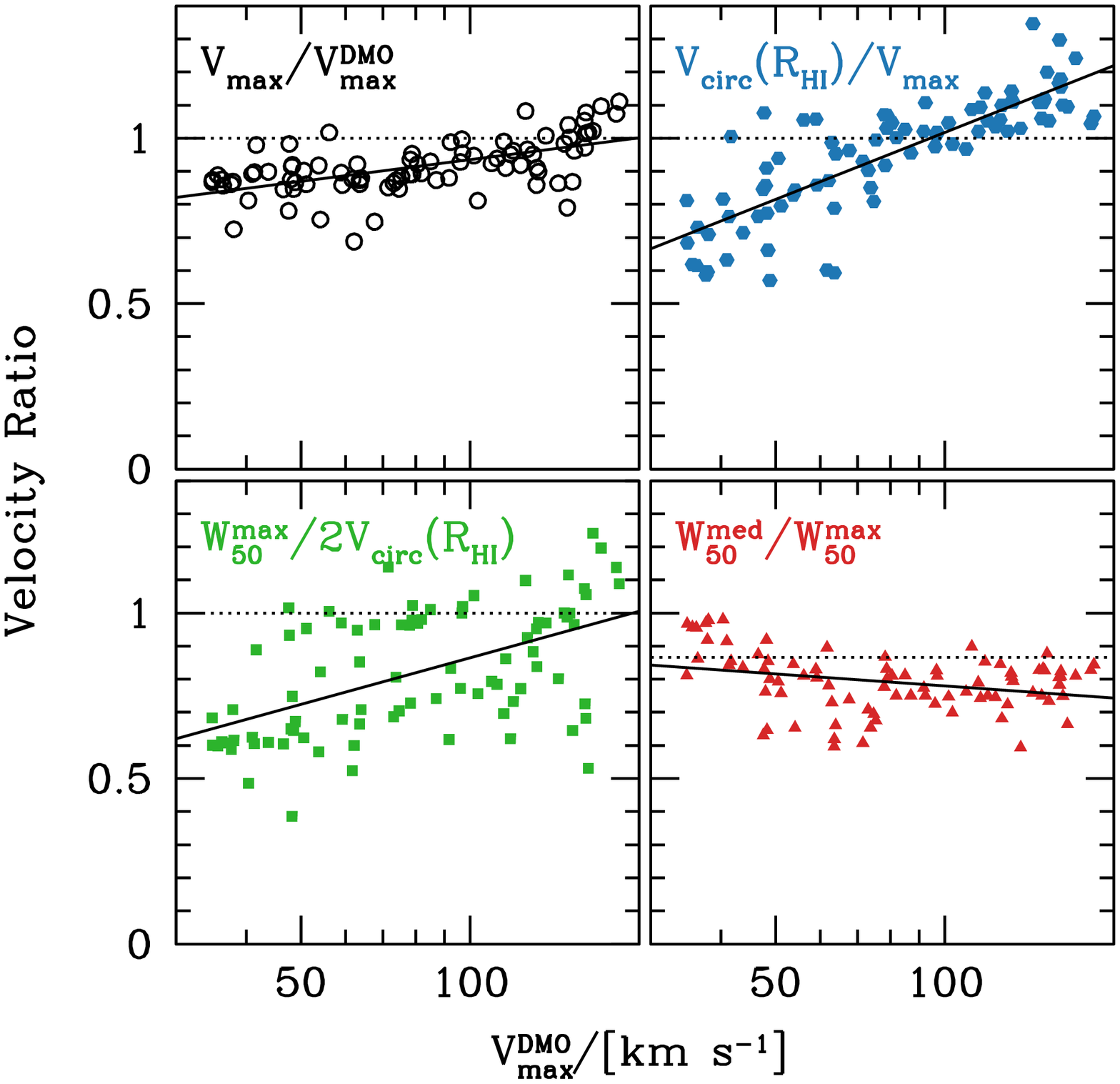}
  \caption{From \VmaxDMO to projected linewidth $W_{50}$ for the full
    sample using four independent ratios.  Upper left shows $V_{\rm
      max}/\VmaxDMO$, the ratio between the circular velocities of the
    hydro and DMO simulation at $R_{\rm max}^{\rm DMO}$. This is
    sensitive to mass in and outflows. The ratio varies from 0.8 to
    1.1. Low velocity haloes have ratios less than unity indicating
    mass loss from the system. Upper right shows $V_{\rm HI}/V_{\rm
      max}$, the ratio between the circular velocity at the HI radius
    and the circular velocity at $R_{\rm max}^{\rm DMO}$. This ratio
    is sensitive to the shape of the circular velocity profile at the
    HI radius.The ratio varies from 0.6 to 1.3. High velocity haloes
    have ratios greater than unity indicating declining velocity
    profiles, while low mass haloes have ratios less than unity
    indicating rising velocity profiles.  The lower left panel shows
    $W_{50}^{\rm max}/2 V_{\rm HI}$, the ratio between the maximum
    half linewidth and the circular velocity. This is sensitive to
    non-circular motions and the shape of the rotation curve. This
    ratio shows the largest variation from 0.4 to 1.2.  The lower right
    panel shows $W_{50}^{\rm med}/W_{50}^{\rm max}$, the ratio between
    the median and maximum linewidths. This ratio is sensitive to the
    effects of projection and indicates whether galaxies behave like
    thin rotating disks (dotted line). The ratio varies from 0.6 to
    1.0.}
\label{fig:vvmax1}
\end{figure*}

Fig.~\ref{fig:vvmax2} shows the relations between various velocity
definitions and $\VmaxDMO$. Linear fits (in $\log-\log$ space) are
given in Table.~\ref{tab:fits}.  Notice that the normalization,
slope, and scatter systematically change at each step. At anyone of
these steps the changes are not large, but they accumulate to a
relation between linewidth and halo velocity that is very different
from a one-to-one correspondence.  For galaxies in Milky Way mass
haloes $\VmaxDMO\simeq 180 \kms$ the average projected half-linewidth
is $W_{50}/2=141\kms$, so that up to the expected project effects, the
linewidth is a good tracer of the maximum halo velocity. However, for
dwarf galaxies at $\VmaxDMO\simeq 50\kms$ the average
half-linewdith of  $W_{50}^{\rm med}/2 \simeq 20\kms$, is a factor of 2.5
times lower than the halo velocity.

There are a number of physical processes that play a role in the
conversion of maximum circular velocity into projected HI linewidths:
inflows and outflows; HI extent and rotation curve shape; non-circular
motions, and projection. These are related, but not uniquely, to  four
ratios, which we show in Fig.~\ref{fig:vvmax1}. 

\subsection{Inflows and Outflows}
The upper left panel shows $\Vmax/\VmaxDMO$, which depends on the net
inflow and outflow of baryons.  At low velocities the ratio $\simeq
0.85$ indicating net outflows, while at high velocities the ratio is
greater than 1, indicating net inflow. As a reference, if a system
lost or did not accrete any of its baryons we would naively expect the
ratio of $\sqrt(1-\fbar)\simeq 0.92$, where the cosmic baryon
fraction, $\fbar=\Omega_{\rm b}/\Omega_{\rm m}\simeq 0.15$.  However,
the reduction in halo velocity is larger than this because the loss of
baryons at early times (as a result of the stellar feedback) reduces
the accretion rate of dark matter, and hence the halo mass and maximum
velocity of the halo by redshift $z=0$.  Similar results are seem in
other cosmological hydrodynamical simulations,
\citet[e.g.,][]{Sawala16}, so the effect appears unrelated to the
details of the sub-grid model for star formation and feedback.

\subsection{HI extent and rotation curve shape}
The upper right panel shows the $\VHI/\Vmax$ ratio vs $\VmaxDMO$.  In
all galaxies $\VHI$ is measured at a smaller radius than $\Vmax$, so
the ratio depends on the shape of the circular velocity profile, and
the extent of the HI gas. In high velocity haloes the ratio is greater
than 1 because the inflow of baryons. In low mass haloes the ratio is
less than 1 because the HI does not extend to the flat part of the
velocity profile. At a halo velocity of 50 km/s the average ratio is
0.8, but there are some dwarf galaxies with a ratio of unity and
others with a ratio of 0.6.  This is already pointing to diversity of
HI extent and/or rotation curve shapes.

\subsection{Non-circular motions}
The lower left panel shows the $W_{50}^{\rm max}/2 \VHI$ ratio. The
linewidth is a convolution of the rotation curve with the HI
distribution. In galaxies that are rotationally supported with a flat
rotation curve this ratio is close to unity. When the ratio is less
than unity this signals a rising rotation curve and/or significant
non-circular motions.  These can be dispersion in the gas, but also
non-axisymmetric features such as bars, spiral arms, and warps, and
out of equilibrium gas flows due to e.g., supernova driven winds or
mergers.  At a halo velocity of 50 km/s the average ratio is 0.7, but
again there is significant scatter, with some dwarf galaxies with a
ratio close to unity.

\subsection{Projection}
The lower right panel shows the ratio between the median and maximum
$W_{50}$ linewidth.  This ratio tells us about projection effects. The
inclination angle, $i$, where $i=0$ is face-on, and $i=90$ is edge-on,
is distributed uniformly in $\cos(i)$ from 0 to 1. For a thin rotating
disk with negligible velocity dispersion, the linewidth at
inclination, $i$, is related to the linewidth at inclination $i=90$ by
$W_i=\sin(i)W_{90}$. The ratio between the median linewidth ($i=60$)
and the maximum ($i=90$) is thus $\sin(60)=0.866$.  In the figure we
see a few galaxies at all halo velocities near this value (dotted
line). However, the majority of galaxies lie below this line
indicating the galaxies are in general not thin rotating disks.  This
is another manifestation of the role of non-circular motions.  At a
halo velocity of 50 $\kms$ the average ratio is 0.8.  At the lowest
velocities the ratio is close to unity, indicating these systems (or
at least the HI lines) have very little rotational support.

\begin{figure}
  \includegraphics[width=0.48\textwidth]{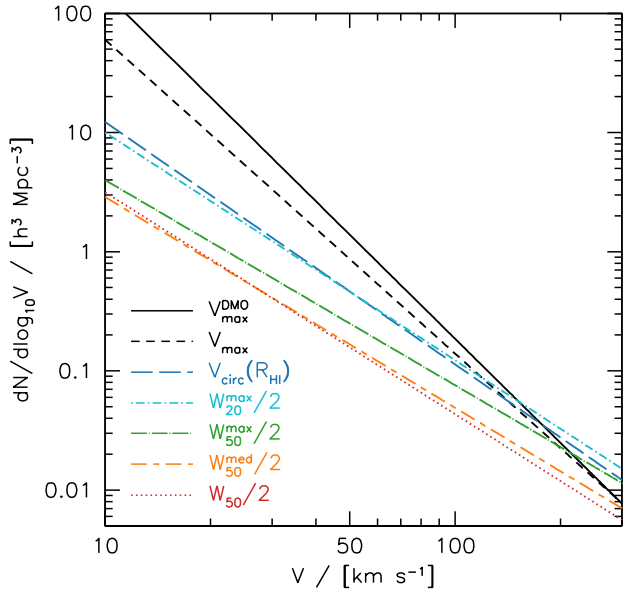}
  \caption{Velocity functions based on different velocity definitions
    calculated using the fitting formula in Fig.3. The slopes get
    progressively shallower as we go from maximum dark halo velocity
    $\VmaxDMO$ to HI half-linewidth $W_{50}/2$.}
    \label{fig:vf1}
\end{figure}

\section{Implied velocity functions}
\label{sec:vf}
We now discuss how the various velocity definitions result in velocity
functions with different slopes.  First we show results with
power-laws, as the transformations are simple and analytic. Then we
consider the impact of Gaussian scatter, and finally the actual
distribution of linewidths from our simulations corresponding to
non-power-laws and non-Gaussian scatter.

\subsection{Analytic considerations for the velocity function}
For a cumulative velocity function with a power-law form
\begin{equation}
  N(>V) = AV^\alpha,
\end{equation}
the differential velocity function has the same slope,
$\alpha$, but with a normalization that is different by the slope:
\begin{equation}
  {\rm d}N/{\rm d}\ln V= \alpha\, N(>V).
\end{equation}
If two different velocities are related by
$V_2=a V_1^b$, then the cumulative velocity function is simply
translated:
\begin{equation}
  N(>V_2) = A (V_2/a)^{\alpha/b}.
\end{equation}
The differential velocity function is changed due to the $d\ln V$, and since
\begin{equation}
  d\ln V_1 = (1/b)\, d\ln V_2,
  \end{equation}
we have the new differential velocity function:
\begin{equation}
  dN/d\ln V_2 = (\alpha/b)\, N(>V_2) = (\alpha/b)\, A (V_2/a)^{\alpha/b}
  \end{equation}

The differential velocity function for $\VmaxDMO$ for Cold Dark Matter
for a Planck cosmology \citep{Planck14} is given by \citep{Klypin15}:
\begin{equation}
  dN/d\log_{10}V = 0.186\,(\VmaxDMO/100)^{-2.9}.
\end{equation}
We can thus use the equations in Table~\ref{tab:fits} to translate
this into the velocity functions of the various velocity
definitions. We first do this ignoring the scatter to show the
magnitude of the effect, later we will include the scatter.  The
result is shown in Fig.~\ref{fig:vf1}, and Table~\ref{tab:vf1} gives
the slopes and normalizations at $V=50\kms$. The majority of the
differences come from four conversions: $\Vmax/\VmaxDMO\simeq 0.63$,
$\VHI/\Vmax\simeq 0.53$, $W_{50}^{\rm max}/2\VHI\simeq 0.54$, and
$W_{50}^{\rm med}/W_{50}^{\rm max}\simeq 0.67$.  The overall shift in
normalization is a factor of 8.7. We stress that this is entirely due
to a change in the velocities, rather than any change in the number
densities of the objects. We note that the strong impact of the
velocity definition on the slope of the velocity function has been
shown previously by  \citet{Brook16a} using halo abundance matching to
link baryonic masses to halo masses, and then various Baryonic
Tully-Fisher relations to relate baryonic masses to velocities. 

\begin{figure}
  \includegraphics[width=0.48\textwidth]{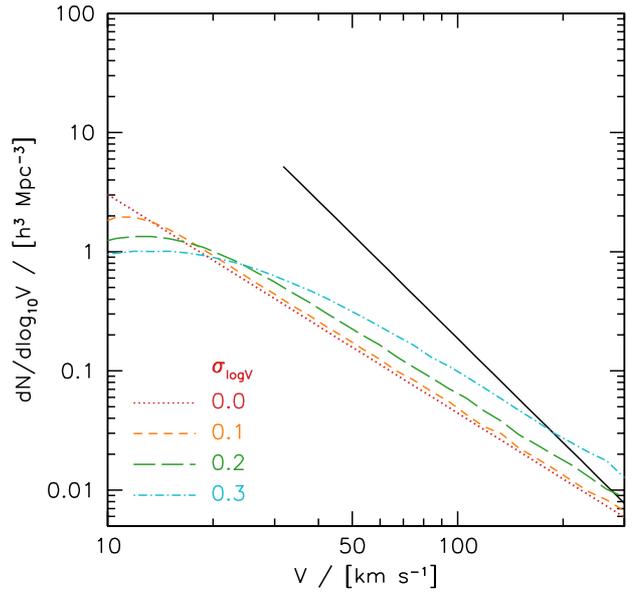}
  \caption{Effect of scatter on the Velocity function. Starting from
    the $W_{50}/2$ velocity function (red dotted line) we add
    log-normal scatter of magnitude 0.1 (orange), 0.2 (green), and 0.3
    (cyan). Low velocity galaxies get preferentially scattered to high
    velocities, thereby increasing the normalization of the velocity
    function. For reference the black line shows the $\VmaxDMO$ function.}
\label{fig:vf2}
\end{figure}

\begin{figure*}
  \includegraphics[width=0.48\textwidth]{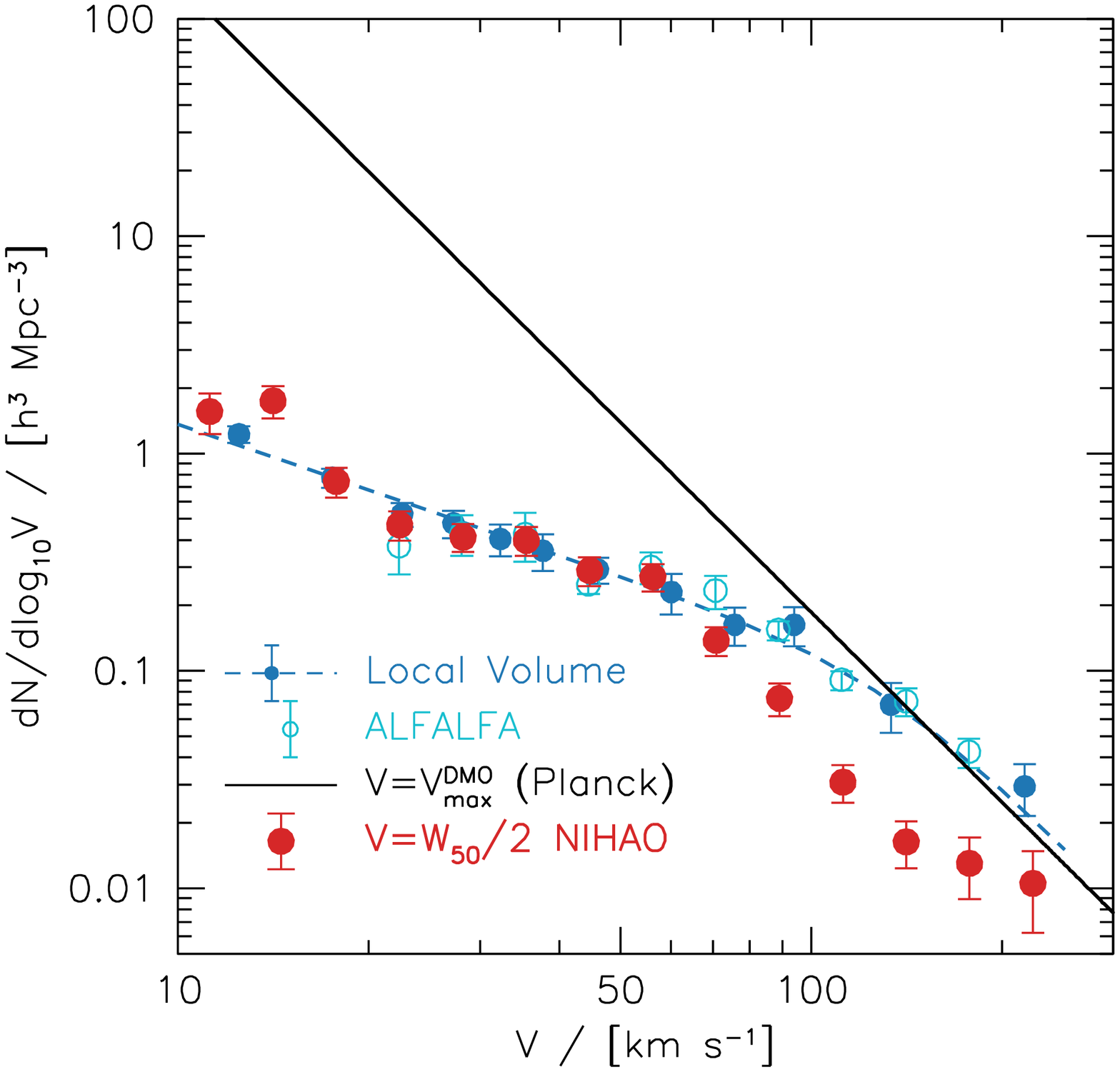}
  \includegraphics[width=0.48\textwidth]{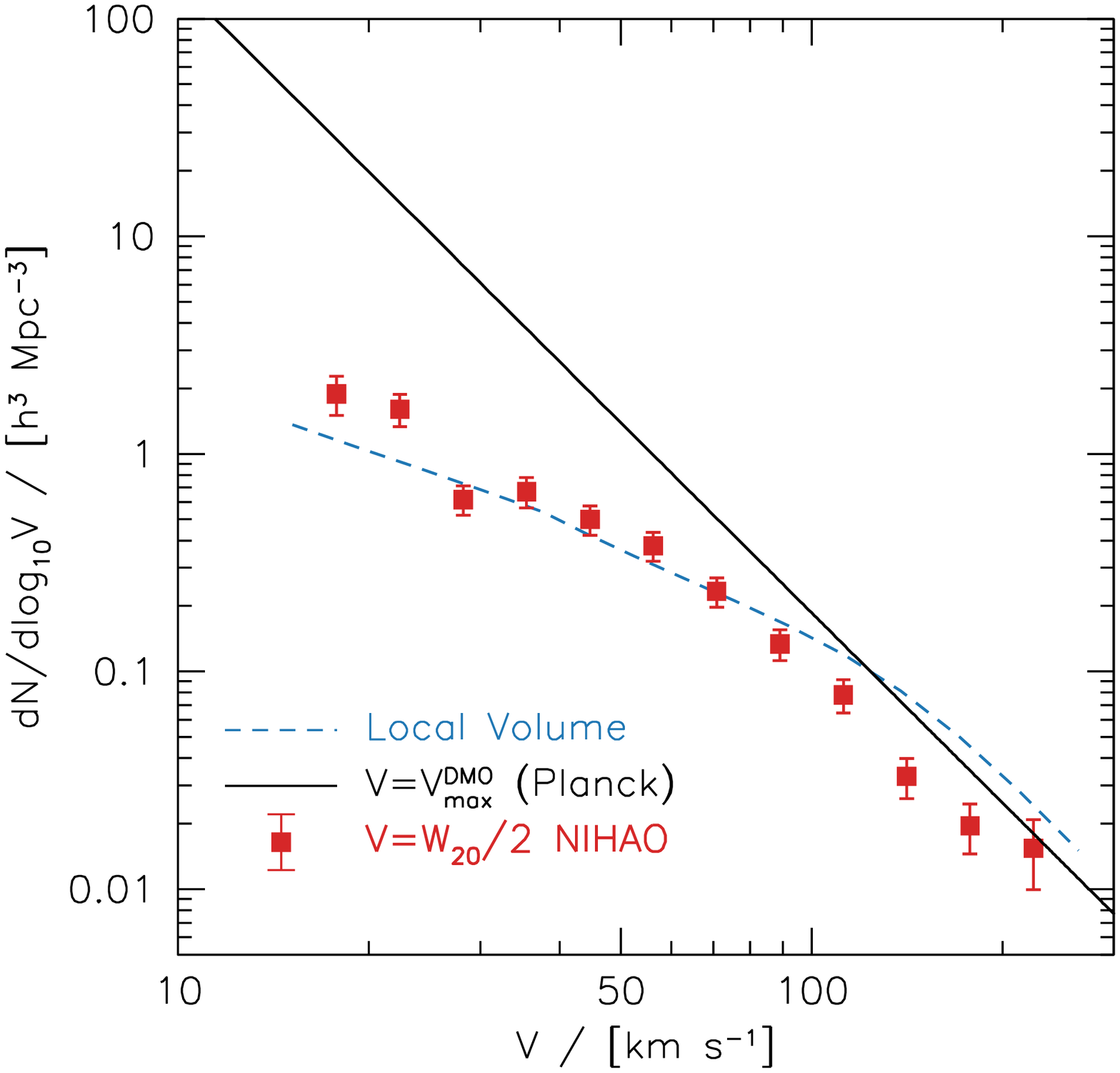}      
  \caption{Linewidth velocity function derived from the full
    distribution of linewidths and $\VmaxDMO$ in the   NIHAO
    simulations (red points) using $W_{50}/2$ (left) and $W_{20}/2$
    (right).  The black solid lines show the CDM halo ($\VmaxDMO$)
    velocity function, which has a steep slope of $-2.9$.
    Observational data from the Local Volume and ALFALFA using
    $W_{50}$ are shown with blue and cyan circles respectively. The blue dashed line shows
    the fit to the Local Volume from \citet{Klypin15}. In the right
    panel we transform the observations using the approximation
    $W_{20}=W_{50}+25 \kms$, with a minimum $W_{50}/W_{20}=0.63$. }
\label{fig:vf4}
\end{figure*}

\begin{table}
\begin{center}
  \caption{Velocity functions for NIHAO simulations from
    Fig.~\ref{fig:vf1} of the form $dN/d\log_{10}V = A
    (V/[50\kms])^{\alpha}$.}
\begin{tabular}{ccc}
\hline
definition & $\alpha$ & $A$ \\
\hline
$V_{\rm max}^{\rm DMO}$ & -2.90 & 1.38 \\
$V_{\rm max}$         & -2.63 & 0.867\\
$V_{\rm HI}$          & -2.03 & 0.464\\
$W_{20}^{\rm max}/2$   & -1.91 & 0.462\\
$W_{50}^{\rm max}/2$   & -1.72 & 0.249\\
$W_{50}^{\rm med}/2$   & -1.77 & 0.166\\
$W_{50}/2$            & -1.80 & 0.158\\
 \hline
\end{tabular}
\label{tab:vf1}
\end{center}
\end{table}

\subsection{Impact of Gaussian scatter}
Fig.~\ref{fig:vf2} shows the impact of Gaussian (in $\log(V)$) scatter
on the velocity function.  The black line shows the velocity function
using $\VmaxDMO$, sampled between a velocity of 30 to 200 $\kms$. The
red dotted line shows the velocity function using $W_{50}/2$. The other
lines show the impact of log-normal scatter (0.1, 0.2, and 0.3 dex) in
the relation between $W_{50}/2$ and $\VmaxDMO$.  The effect of scatter
is to increase the normalization of the velocity function.  Galaxies
get preferentially scattered to higher velocity because there are many
more low velocity haloes than high velocity haloes. The overall effect
is not large, just 0.1 dex change for a scatter of 0.2 dex.

\begin{table}
\begin{center}
  \caption{Differential velocity functions for NIHAO simulations from Fig.~\ref{fig:vf4}.}
\begin{tabular}{ccccc}
\hline
$\log(V)$ & $dN/d\log(V)$ & error & $dN/d\log(V)$ & error \\
            & $W_{50}/2$ &  & $W_{20}/2$ &  \\
 \hline
1.05 & 1.559  & 0.332   & --     & --\\
1.15 & 1.747  & 0.295   & --     & --\\
1.25 & 0.743  & 0.116   & 1.888  & 0.385\\
1.35 & 0.469  & 0.072  & 1.603  & 0.271\\
1.45 & 0.413  & 0.061  & 0.617  & 0.095\\
1.55 & 0.398  & 0.061  & 0.671  & 0.107\\
1.65 & 0.289  & 0.044  & 0.500  & 0.078\\
1.75 & 0.270  & 0.039  & 0.378  & 0.058\\
1.85 & 0.138  & 0.021  & 0.233  & 0.036\\
1.95 & 0.0747 & 0.0128  & 0.134  & 0.021\\
2.05 & 0.0308 & 0.0060 & 0.0781 & 0.0136\\
2.15 & 0.0163 & 0.0040 & 0.0330 & 0.0069\\
2.25 & 0.0130 & 0.0041 & 0.0195 & 0.0050\\
2.35 & 0.0105 & 0.0043 & 0.0154 & 0.0054\\
 \hline
\end{tabular}
\label{tab:vf2}
\end{center}
\end{table}

\subsection{Deriving the velocity function from linewidth vs maximum circular velocity}

We now go into more detail and show how the distribution of galaxies
in the $W_{50}/2$ vs $\VmaxDMO$ plane can be converted into the
velocity function.  We set up a grid in $x=\log(\VmaxDMO)$ and $y=\log(W_{50}/2)$ of width $dx$ and
$dy$ and count the number of galaxies in each cell, $n(x,y)$.  From
the cumulative DMO velocity function we straightforwardly know the
number of dark matter haloes in a given $x$-bin in a given volume of
space, $dN_{\rm CDM}(x)$. Comparing the actual number of data points
$N(x)$ (number of haloes in each $x$-bin multiplied by number of
projections per halo) to $dN_{\rm CDM}(x)$ we get the weight $w(x)$
for each halo in bin $x$:
\begin{equation}
  w(x)=\frac{dN_{\rm CDM}(x)}{N(x)}
\end{equation}
We then go to each $y$-bin and count up the
number of galaxies using the weights:
\begin{equation}
  N(y)=\frac{\Sigma_{x} n(x,y) w(x)} {\Sigma_{x} w(x)}.
\end{equation}

As fiducial bin widths we adopt $dx=0.05$ and $dy=0.10$.  The
resulting velocity functions for $W_{50}$ (left) and $W_{20}$ (right)
are shown with red points in Fig.~\ref{fig:vf4}, and tabulated in
Table~\ref{tab:vf2}. The error bars ($e_y$) are calculated as
$e_y=y/\sqrt{N}$, where $N$ is the number of simulations that
contribute to each $y$-bin.  At low velocities ($W_{50}/2 < 80 \kms$)
the NIHAO $W_{50}/2$ function has a normalization and shallow slope
$\sim-1$ in agreement with observations from the Local Volume
\citep{Klypin15} and ALFALFA \citep{Papastergis16}.  At high
velocities ($\gta 100 \kms$) the NIHAO simulations underpredict the
observed number densities, or alternatively the linewidths are too
low.  We return to a possible cause of this in section
~\ref{sec:lineshape} below.

\begin{figure*} 
  \includegraphics[width=0.45\textwidth]{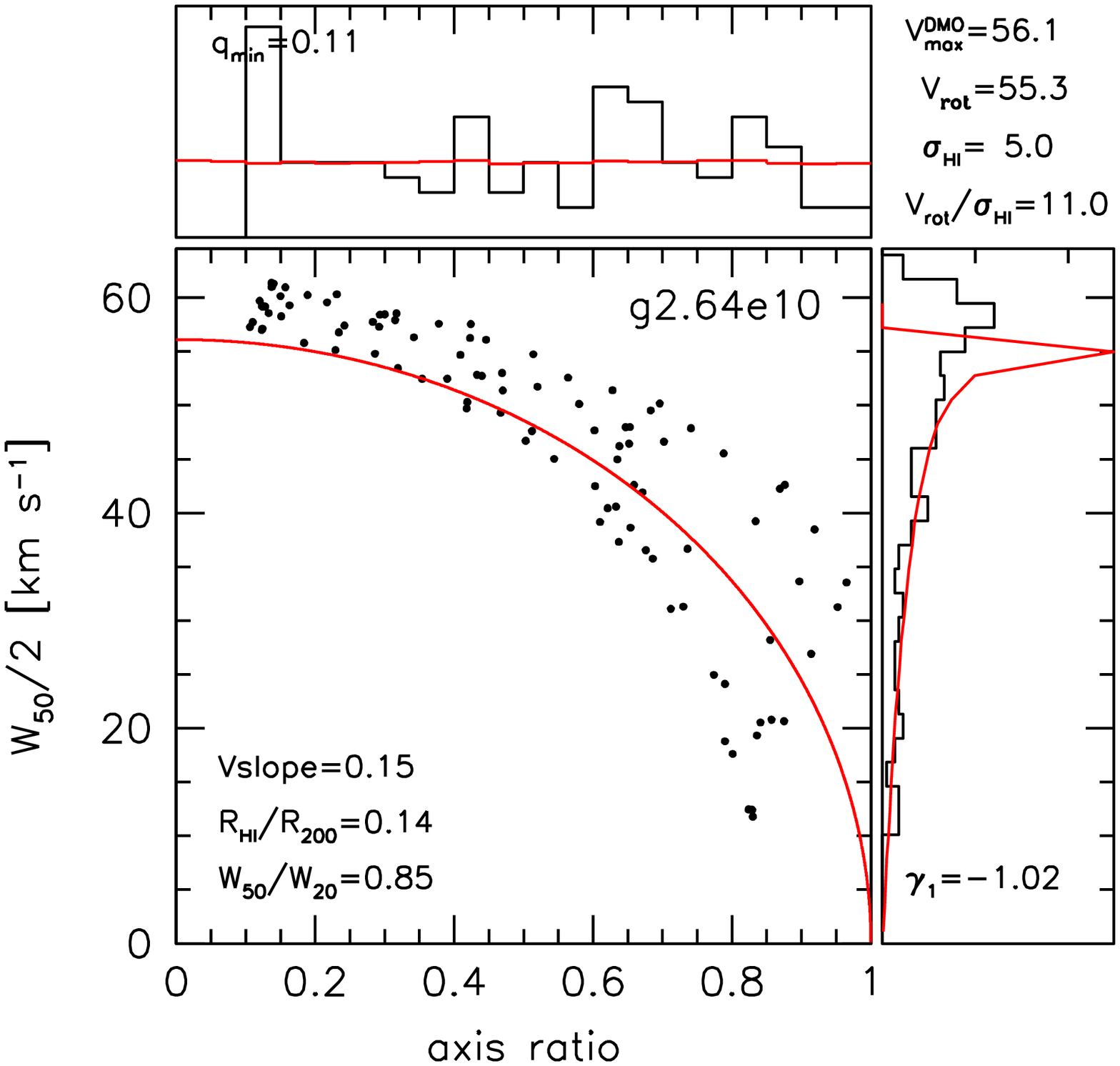}
  \includegraphics[width=0.45\textwidth]{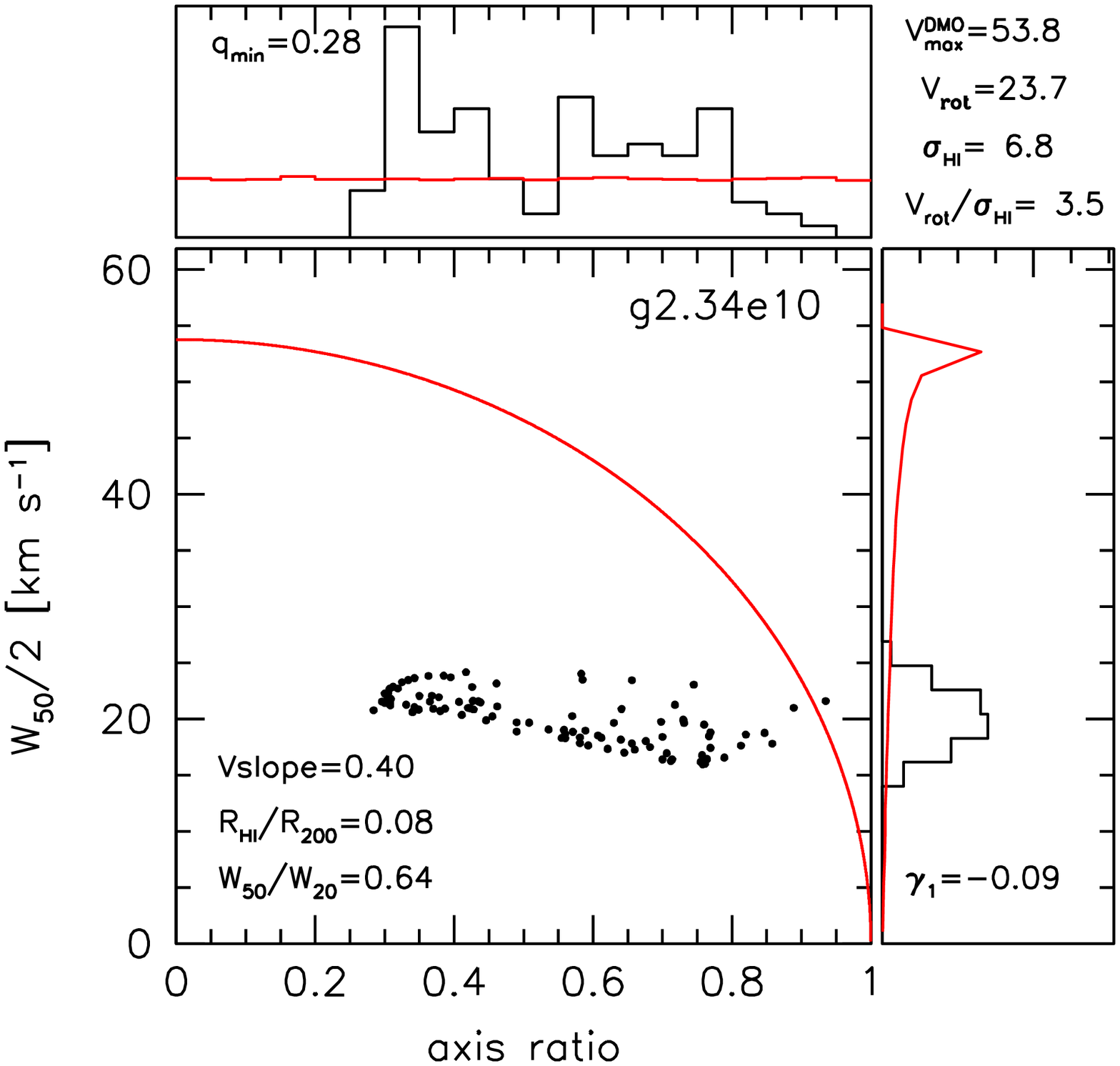}
  \includegraphics[width=0.45\textwidth]{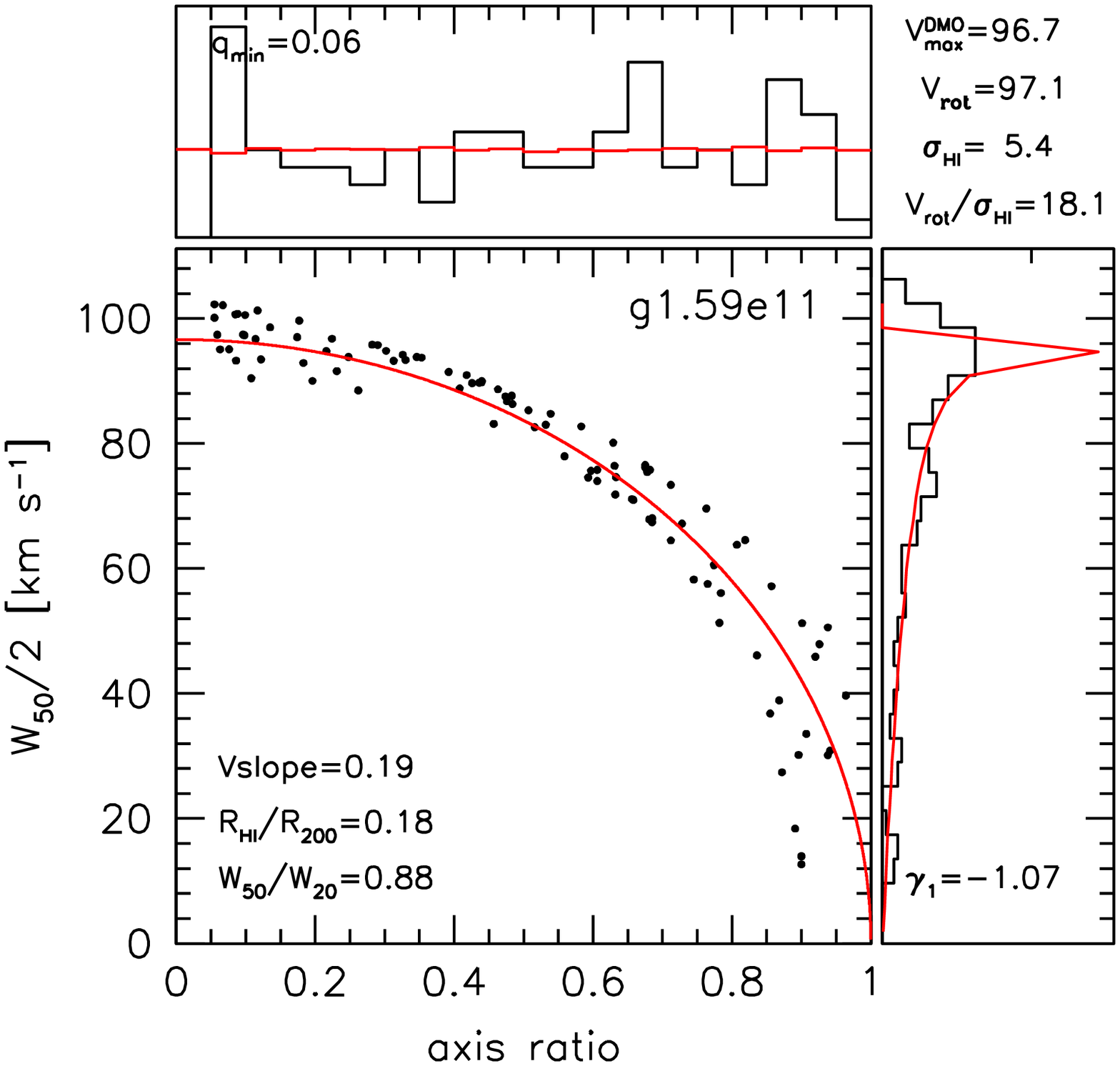}
  \includegraphics[width=0.45\textwidth]{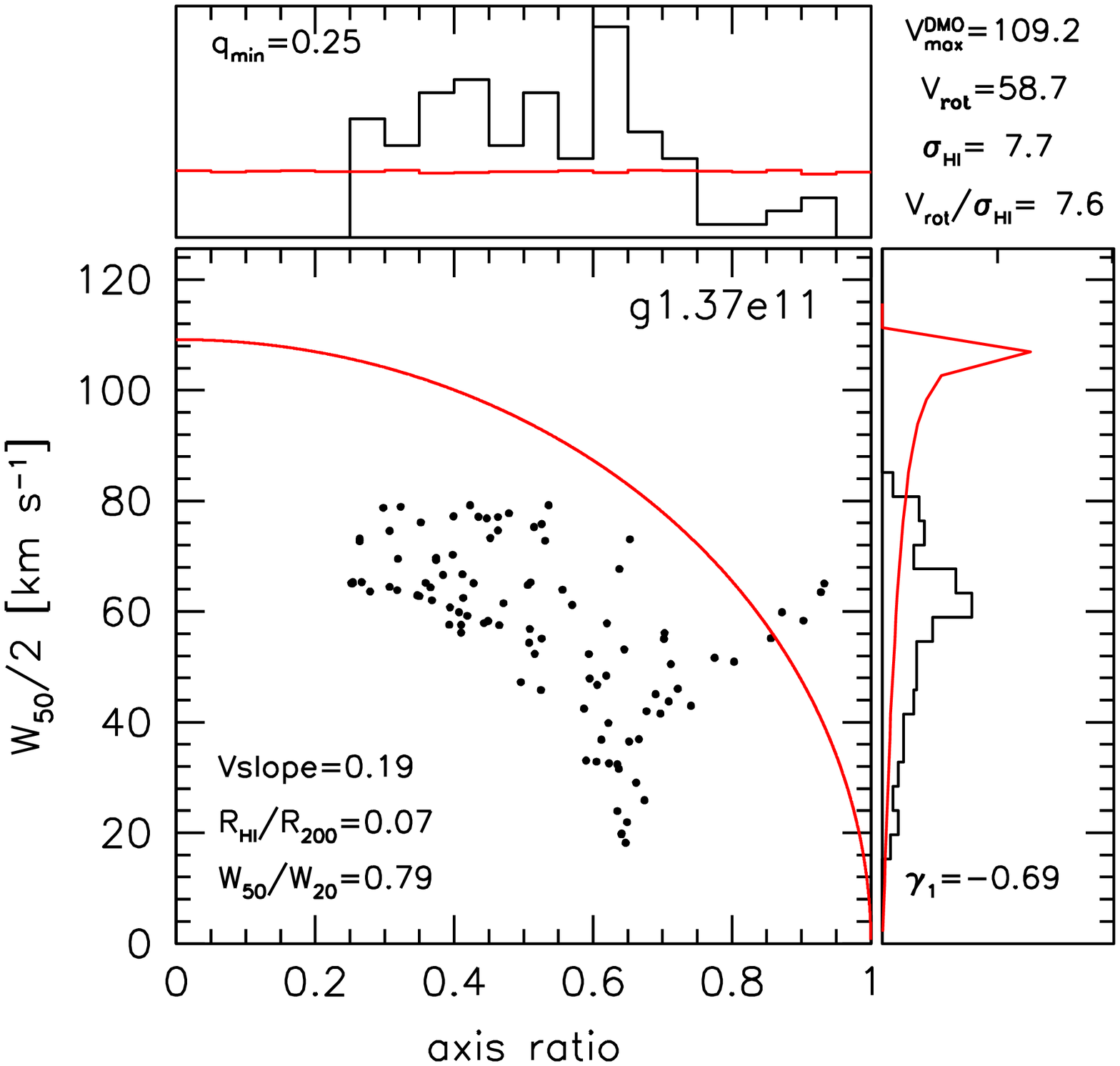}
  \caption{Projection effects in the linewidth vs axis ratio plane for
    four test galaxies. Upper panels show galaxies with maximum
    circular velocity $\VmaxDMO\sim 55\,\kms$, while lower panels show
    galaxies with $\VmaxDMO \sim 100\,\kms$. Left panels show rotation
    dominated galaxies, right panels show galaxies with more pressure
    support, as can be seen by the thicker HI disks ($q_{\rm min}$),
    and more symmetric distributions of linewidths ($\gamma_1$). The
    red lines show the prediction for a randomly oriented thin disk
    rotating at ($\VmaxDMO$) -- uniform axis ratios, and linewidths
    scaling as $\sin(i)$, where $\cos(i)$ is uniformly distributed.}
\label{fig:projection}
\end{figure*}

In our simulations $W_{20}/2$ is a better predictor of the halo
velocity than $W_{50}/2$ (See Fig.~\ref{fig:vvmax2}) so this is our
preferred definition for future observations.  Currently observations
of the $W_{20}$ function are not available, so we shift the
observations using the approximation $W_{20}=W_{50}+25 \kms$
\citep{Brook16b}, and a minimum of $W_{50}/W_{20}=0.63$.
In this case the simulations are closer to 
the observed velocity function at high velocities, while
maintaining the agreement at low velocities.

\begin{figure*} 
  \includegraphics[width=0.90\textwidth]{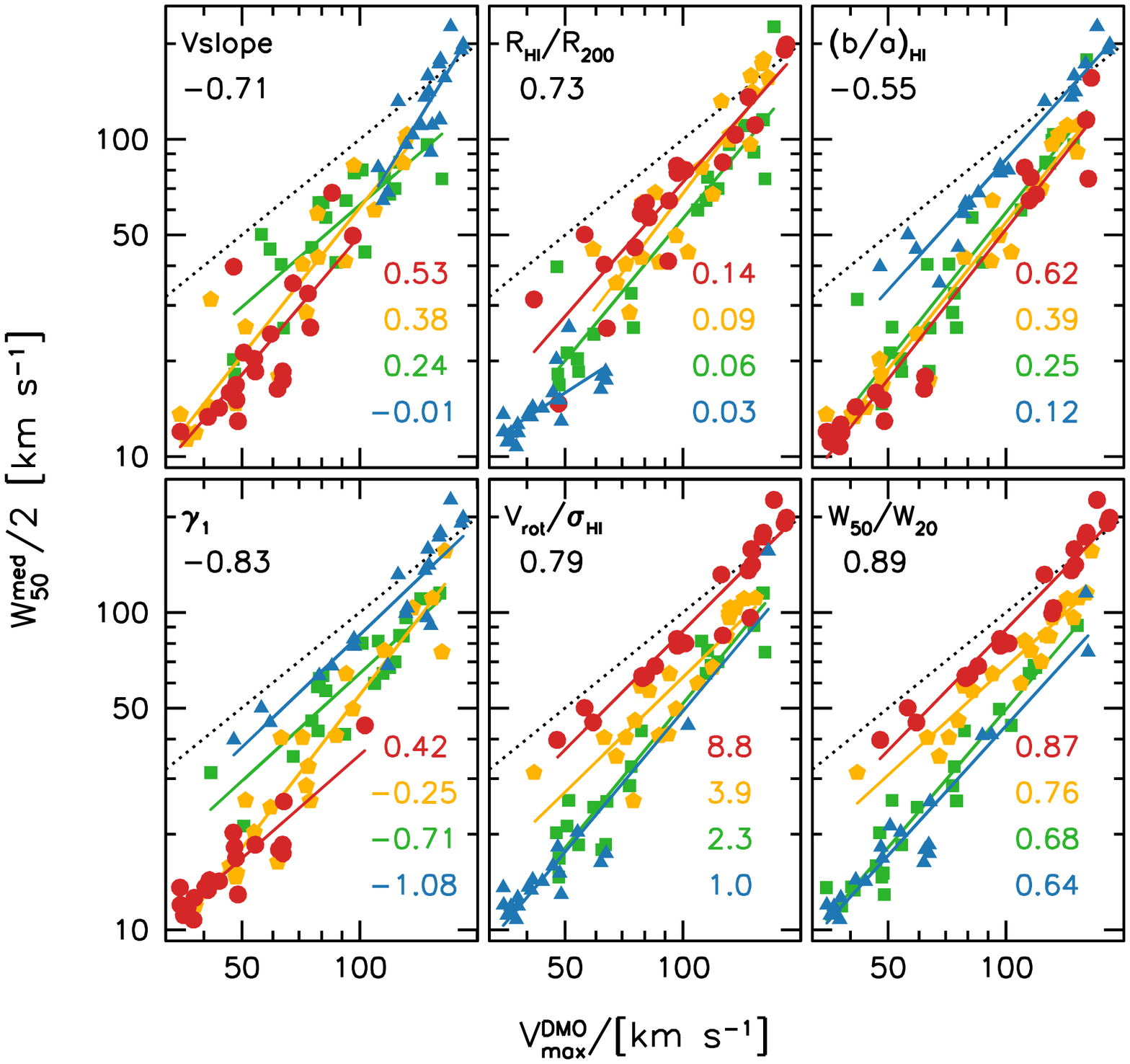}
  \caption{Relation between median HI linewidth and maximum circular
    velocity of the dark matter halo. Points are color coded by the
    parameter indicated in the top left corner of each panel. The
    number at the top left corner is the correlation coefficient
    between said parameter and $W_{50}^{med}/2$ and $\VmaxDMO$.
    Structural parameters: logarithmic slope of the outer circular
    velocity profile (top left);  HI-to-virial radius (top middle);
    minimum minor-to-major HI axis ratio (top right). Kinematic
    parameters: skewness of the linewidth distribution (bottom left);
    Rotation-to-dispersion ratio (bottom middle); and linewidth
    profile shape (bottom right).  Each color corresponds to a
    quartile of the distribution (red are the largest 25\%, while blue
    are the smallest 25\%), the mean of each quartile is indicated by
    the colored numbers.}
\label{fig:wvmax2}
\end{figure*}

\section{Testing the simulations}
\label{sec:tests}
We have shown that the discrepancy at low velocities between the
observed HI linewidth function and the $\VmaxDMO$ function of LCDM is
resolved by the NIHAO simulations. We now discuss in more detail why
this is the case, and observational ways in which the simulations can
be tested. We focus our discussion on six (mostly dimensionless)
parameters that are correlated with the variation in HI linewidth at
fixed maximum dark halo velocity.

\begin{itemize}
\item Rotation-to-dispersion ratio, $V_{\rm rot}/\sigma_{\rm HI}$. The
  HI dispersion is calculated from the minimum HI linewidth,
  $\sigma_{\rm HI}=W_{50}^{\rm min}/2.35$.  For the rotation velocity
  we use the maximum of the rotation curve, which tends to occur near
  the HI radius (see Fig.~\ref{fig:vr} for examples). This definition is not
  directly observable, but similar definitions are.
\item $\gamma_1$, Skewness of the $W_{50}$ distribution, see below for
  definition.  Can be measured on samples of galaxies.
\item HI line profile shape, $W_{50}/W_{20}$. Directly observable.
\item  Outer circular velocity curve slope, $\Delta \log V /\Delta
  \log R$. For our simulations is measured between 0.5 and 1.0 $R_{\rm
    HI}$. Can be measured for individual galaxies. 
\item HI-to-virial radius, $R_{\rm HI}/R_{200}$. Dimensionless, but
  not directly observable. HI sizes are observable.
\item HI disk thickness. Computed as the minimum HI axis ratio from
  all projections. Can be measured from samples of galaxies.
\end{itemize}

\begin{figure*} 
  \includegraphics[width=0.90\textwidth]{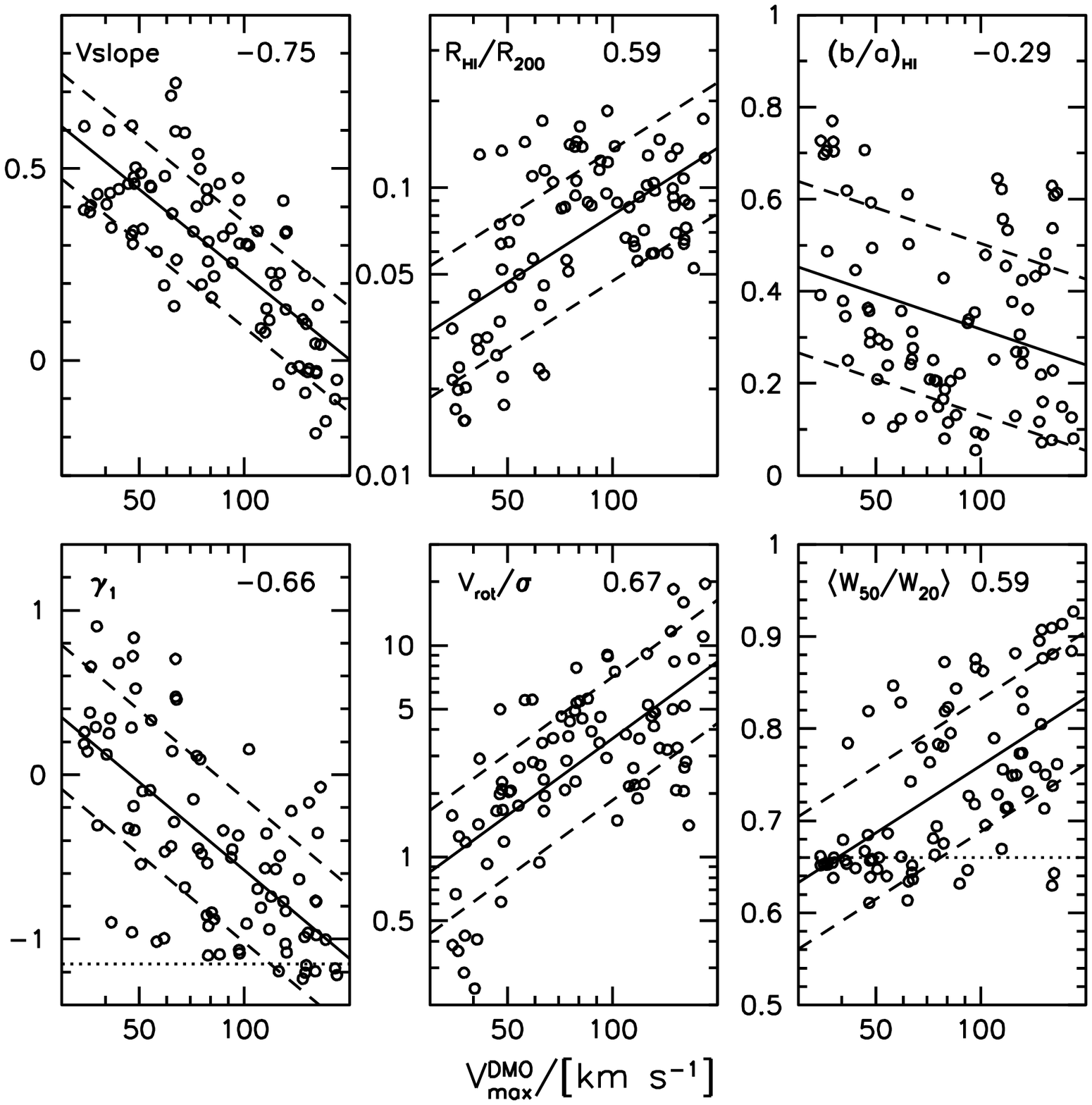}
  \caption{Parameters vs halo velocity. The parameter in question is
    indicated in the top left corner. The correlation coefficient is
    given in the top right corner. The mean and standard deviation of
    a linear fit is shown with sold and dashed lines. Dotted
    horizontal lines indicate the expectation for a rotating disk
    (lower left) and Gaussian (lower right).}
\label{fig:vsvmax}
\end{figure*}

\subsection{Testing rotational support with projection effects}

One of the key results from our simulations is that galaxies, and
dwarf galaxies in particular, do not always have HI kinematics as
expected from idealized thin rotating disks. An observationally
testable consequence of this is the effect of projection.

For a thin rotationally supported axisymmetric disk the minor-to-major
axis ratio $(b/a)$ is uniformly distributed between 0 and 1. In terms
of the disk inclination, $i$, where $i=0$ is face-on and $i=90$ is
edge-on, $b/a=\cos(i)$, the observed linewidth then varies as
$\sin(i)$. Fig.~\ref{fig:projection} shows the linewidth vs axis ratio
for our four test galaxies. The axis ratio is computed for each
projection using moments as described in section 2.  The upper
galaxies have maximum halo velocity $\VmaxDMO\simeq 55 \kms$, while
the lower two galaxies have $\VmaxDMO\sim 100 \kms$. The corresponding
stellar masses are $\sim 2\times 10^8$ and $\sim 10^9\Msun$,
respectively.

The two galaxies on the left panels of Fig.~\ref{fig:projection} have
close to uniform axis ratio distributions (upper histograms) and
strongly skewed linewidth distributions (right histograms) close to
that predicted for thin disks rotating at the maximum velocity of the
dark matter halo (solid red lines).  On the other hand, the two
galaxies in the right panels of Fig.~\ref{fig:projection} strongly
deviate from this prediction with a lower mean linewidths and a more
symmetric distribution of linewidths. The axis ratio distribution is
also more centrally concentrated.

A parameter that is related to how rotationally supported a
galaxy is, is the skewness of the linewidth distribution. Here we
adopt Pearson's moment coefficient of skewness, which 
is defined as
\begin{equation}
  \gamma_1 = \mu_3 / \sigma^3.
\end{equation}
Here $\mu_3=\Sigma (W-\overline{W})^3 / N$ is the third central
moment, $\sigma$ is the standard deviation, and $\overline{W}$ is the
mean. A value of $\gamma_1=0$ corresponds to a symmetric distribution,
$\gamma_1<0$ corresponds to a tail to low values, while $\gamma_1>0$
corresponds to a tail to high values. A thin rotating disk has a
strongly negatively skewed distribution of linewidths with
$\gamma_1=-1.13$.

The galaxies in the left panels of Fig.~\ref{fig:projection}  have
$\gamma_1=-0.97$ and $-1.06$, i.e., they behave like thin rotating
disks. In contrast the galaxies on the right have $\gamma_1=-0.10$,
and $\gamma_1=-0.73$, and are thus less rotationally supported
systems.  These inferences agree with the actual rotation to
dispersion ratios ($V_{\rm rot}/\sigma_{\rm HI}$). 

Another  parameter that is related to the rotational support of a
galaxy is minimum minor-to-major axis ratio, $q_{\rm min}$ (i.e., the
disk thickness). The galaxies on the left have smaller minimum axis
ratios (corresponding to thinner edge-on disks): $q_{\rm min}=0.11$
and $0.06$ compared to $0.28$ and $0.25$.

The ratio between linewidths $W_{50}/W_{20}$, also depends on the
rotational support, as well as the circular velocity curve shape, and
the extent of the HI.  The galaxies on the left have steeper line
profiles, and more extended and shallower outer circular velocity
curve slopes.

\subsection{What causes linewidths to underestimate maximum circular velocity?}

Fig.~\ref{fig:wvmax2} shows the impact of several (mostly
dimensionless) parameters on the relation between median HI linewidth
and maximum dark halo circular velocity. The points are color coded by
the parameter in question.  The four numbers in each panel indicate
the mean value of each quartile, the solid line is a fit to each
quartile of points. We discuss each of them in turn below.

\begin{itemize}
\item Outer circular velocity curve slope (top left). A positive value
  indicates that the curve is still rising in the outer parts, while a
  value close to zero indicates a flat velocity curve.  Galaxies with
  more rising circular velocity profiles (red points) tend to have
  lower linewidth-to-halo velocity ratios. 

\item HI-to-virial radius (top middle). Smaller galaxies (blue
  triangles) have lower linewidths.  At least partially because if the
  HI does not reach the flat part of the circular velocity curve, then
  the kinematics will progressively underestimate the maximum circular
  velocity. 

\item HI axis ratio (top right). The thinnest galaxies
  (red points, $b/a\sim 0.1$) have linewidths that trace the halo
  velocity, while thicker galaxies have progressively lower
  linewidths.
  
\item $\gamma_1$, Skewness of the $W_{50}$ distribution (top
  left). Galaxies with strong negative skew (blue points) have
  linewidths that trace the maximum halo velocity. Larger values of
  $\gamma_1$ result in lower linewidths.
  
\item Rotation-to-dispersion ratio (bottom middle).  Rotationally
  supported systems (red circles) tend to have linewidths that trace
  the halo velocity, while pressure supported systems (blue triangles)
  tend to have low linewidths.

\item HI line profile shape (bottom left). Steeper line profiles
  $W_{50}/W_{20}\sim 0.9$ (red points) have linewidths close to halo
  velocity, while shallow line profiles $W_{50}/W_{20}\sim 0.6$ (blue
  triangles) have linewidths significantly below the halo velocity. The
  profile shape depends on a combination of the rotation curve shape and
  extent, and the rotation to dispersion ratio.

\end{itemize}

The number in the top left indicates the correlation coefficient
between the specific parameter and $W_{50}^{\rm med}/2 \VmaxDMO$.
Line profile shape shows the strongest correlation ($\rho=0.89$)
closely followed by the skewness ($\rho=-0.83$), and HI rotation to
dispersion ratio  ($\rho=0.79$).  At a given halo velocity all six
parameter show a correlation with linewidth. To see which of these
parameters is most likely to explain the trend with halo mass we  show
the dependence on halo velocity in Fig.~\ref{fig:vsvmax}. The straight
and dashed lines show a fit with 1$\sigma$ scatter. The number gives
the correlation coefficient. We see clear trends but also significant
scatter.  Structurally, we see that lower velocity haloes have steeper
outer circular velocity curve slopes and relatively smaller HI sizes,
and marginally thicker HI disks.  Kinematically, we see that lower
velocity haloes have less negatively skewed HI linewidth
distributions, lower rotation-to-dispersion ratios, and shallower line
profiles.

In summary we see that the dependence of $W_{50}$ with $\VmaxDMO$ is
driven primarily by two effects: The degree of rotational support, and
the shape of the rotation curve. Galaxies in lower mass halos have
less rotational support, and less extended HI. Below we show current
observations that show evidence of these two effects.

\begin{figure*} 
  \includegraphics[width=0.96\textwidth]{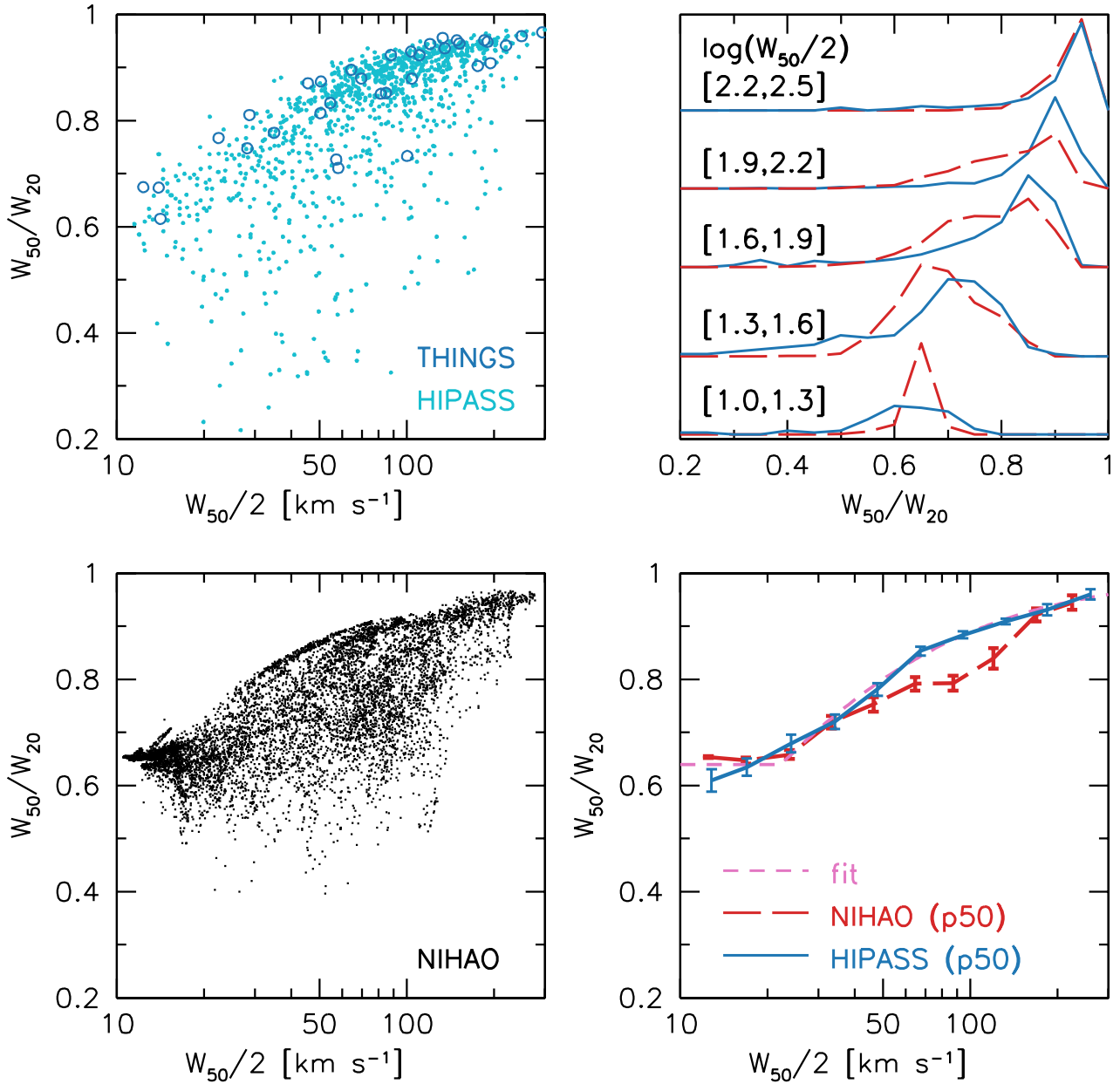}
  \caption{Linewidth ratio $W_{50}/W_{20}$ vs $W_{50}/2$. NIHAO
    simulations are shown with black points (100 projections per
    galaxy) and red long-dashed lines. HIPASS observations are shown
    with cyan points and blue solid lines. THINGS observations are
    shown with blue circles. The left panels show the distribution of
    points, the lower right panel shows the median relations, where
    the error bars show the error on the median. For NIHAO this is the
    error corresponding to the number of distinct galaxies in each
    bin, rather than the number of data points (which is $\sim 30$
    times higher). The dashed magenta line is a simple fit to the
    observations: $W_{20}=W_{50} +25 \,\kms$ with a minimum of
    0.64. The upper right panel shows the distribution of
    $W_{50}/W_{20}$ in five bins of $W_{50}/2$ as indicated.}  
    \label{fig:ww}
\end{figure*}

\begin{figure*} 
  \includegraphics[width=0.90\textwidth]{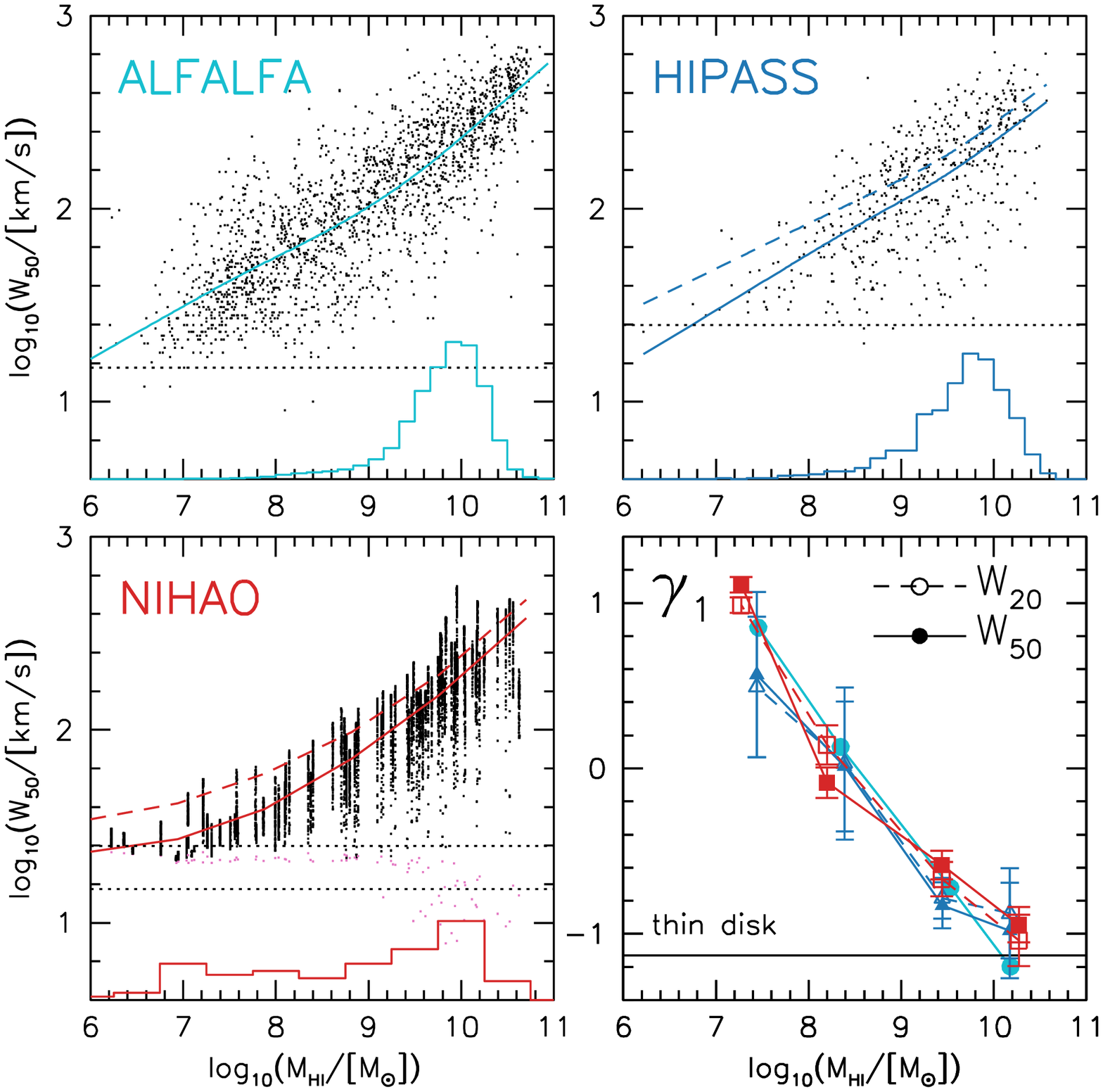}
  \caption{The HI linewidth - mass relation. Upper panels show the
    observed relations from ALFALFA and HIPASS. Points are $W_{50}$,
    solid lines show the mean relation. Dashed lines show the mean
    relation using $W_{20}$. Only 100 points per 0.2 dex in mass are
    plotted to make the change in scatter with mass more apparent. The
    histogram shows the distribution of mass for the full sample. The
    horizontal dotted lines show the instrumental velocity resolution
    limit.  The lower left panel shows the relation from NIHAO, where
    we plot 50 random projections per simulated galaxy. The magenta
    points show the linewidth due to thermal broadening.  The lower
    right panel shows the skewness of the distribution in bins of
    mass. Error bars are from bootstrap resampling. An idealized thin
    rotating disk has $\gamma_1=-1.13$ (horizontal black line). High
    mass galaxies in both observations and simulations are close to
    this value. While lower mass galaxies progressively deviate
    suggesting more disordered kinematics.}
\label{fig:wmass}
\end{figure*}

\subsection{Line profile shape}
\label{sec:lineshape}
The shape of the HI line profile depends on the amount of rotational
support and the shape of the rotation curve.  A simple way to
parameterize this is the ratio between  linewidths measured at the
50\% and 20\% level of peak flux, $W_{50}$ and $W_{20}$. The ratio is
close to 1 for a rotating disk with a flat rotation curve whose
velocity dispersion is small compared to the line-of-sight component
of the rotation velocity.  As the outer rotation curve slope becomes
more positive the ratio decreases, since a larger fraction of the HI
flux is coming from gas rotating lower than the maximum. Pressure
support also causes ratio to decrease. In the limit $V/\sigma=0$ the
line profile is simply that due to random motions of the gas, and is
independent of the rotation curve shape.  For a Gaussian
$W_{50}/W_{20}=\sqrt{\ln(0.5)/\ln(0.2)}\simeq 0.66$.  The
observational advantage of this ratio is it does not require spatially
resolved HI observations, which are currently not feasible for large
samples of galaxies.

A comparison between NIHAO simulations and observations was previously
shown in \citep{Maccio16}. In both observations and simulations the
$W_{50}/W_{20}$ decreases for lower linewidth galaxies. This result
was also shown by \citep{Brook16b} using a smaller sample of galaxies
from the MaGICC project (the precursor to NIHAO), and by
\citet{El-Badry18b} using galaxies from the FIRE project.

Fig.~\ref{fig:ww} shows the relation between $W_{50}$ and $W_{20}$
from NIHAO simulations compared to observations from HIPASS
\citep{Koribalski04} and THINGS \citep{Walter08}.  In both simulations
and observations the ratio varies from  $\sim 0.95$ at high velocities
to $\sim 0.65$ at low velocities, and the amount of scatter is also
similar.  However, when looking at the median relations (lower right
panel) the NIHAO simulations underpredict the HIPASS observations for
$W_{50}\sim 100 \kms$. This is the same scale where the simulations
underpredict the observed velocity function (Fig.~\ref{fig:vf4}).  The
upper right panel shows the distributions of $W_{50}/W_{20}$ in five
bins of $W_{50}/2$. Again this shows an excess of galaxies in NIHAO
with low linewidth ratios at intermediate velocities. In principle
this could be due to there being too much non-circular motions and/or
the HI disks being too small at this velocity scale. Since the NIHAO
simulations are in good agreement with observations of HI sizes (see
below), we think the low linewidth ratio is more likely due to too an
excess of non-circular motions, likely driven by feedback. Due to the
large intrinsic variation in $W_{50}/W_{20}$ and the relatively small
sample size, another contributing cause could be sampling effects,
i.e., we by chance happen to simulate more galaxies with lower
ratios. A similar discrepancy in $W_{50}/W_{20}$ at this velocity
scale is seen in the FIRE simulations \citep{El-Badry18b} and likely
has a common origin, which points to feedback driven turbulence.

In summary, the \hi line profile shape is a powerful, yet simple test
of the simulations, and thus motivates future blind \hi surveys
obtaining deep enough data to accurately measure both $W_{50}$ and
$W_{20}$.

\begin{figure} 
  \includegraphics[width=0.45\textwidth]{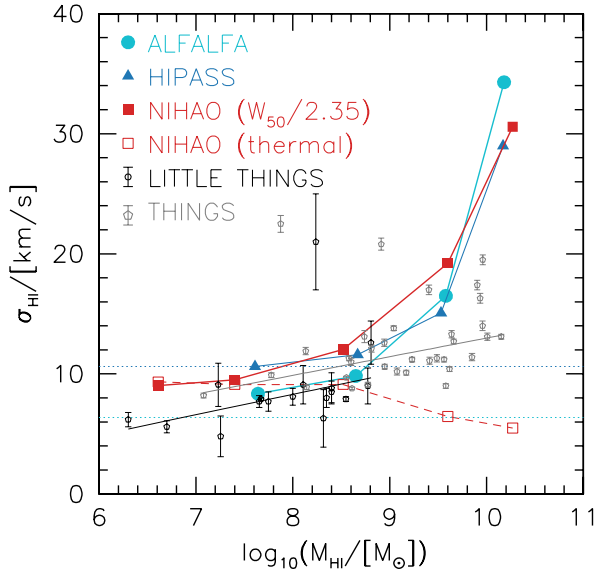}
  \caption{\hi velocity dispersion vs \hi mass. For NIHAO (red),
    ALFALFA (cyan), and HIPASS (blue) the velocity dispersion is
    estimated from the ``face-on'' linewidths using $\sigma_{\rm
      HI}=\langle W_{50}\rangle/2.35$, where $\langle W_{50}\rangle$
    is the average of the lowest 5\% of linewidths in a given bin of
    HI mass.  For thin rotating disks, this includes a non-neglible
    contribution from rotation, so should be considered an upper limit
    to $\sigma_{\rm HI}$. For NIHAO the open squares show the average
    of the thermal velocity dispersion. The horizontal dotted lines
    show the resolution limits of ALFALFA (cyan) and HIPASS (blue),
    and suggest that the HIPASS dispersions are effected by the
    resolution limit. Measurements for individual galaxies using
    resolved \hi observations from LITTLE THINGS \citep{Iorio17} and
    THINGS \citep{Ianja12} are shown with black and grey points with
    error bars, respectively. The lines show fits to these data. } 
\label{fig:sigmaHI}
\end{figure}

\begin{figure*} 
  \includegraphics[width=0.96\textwidth]{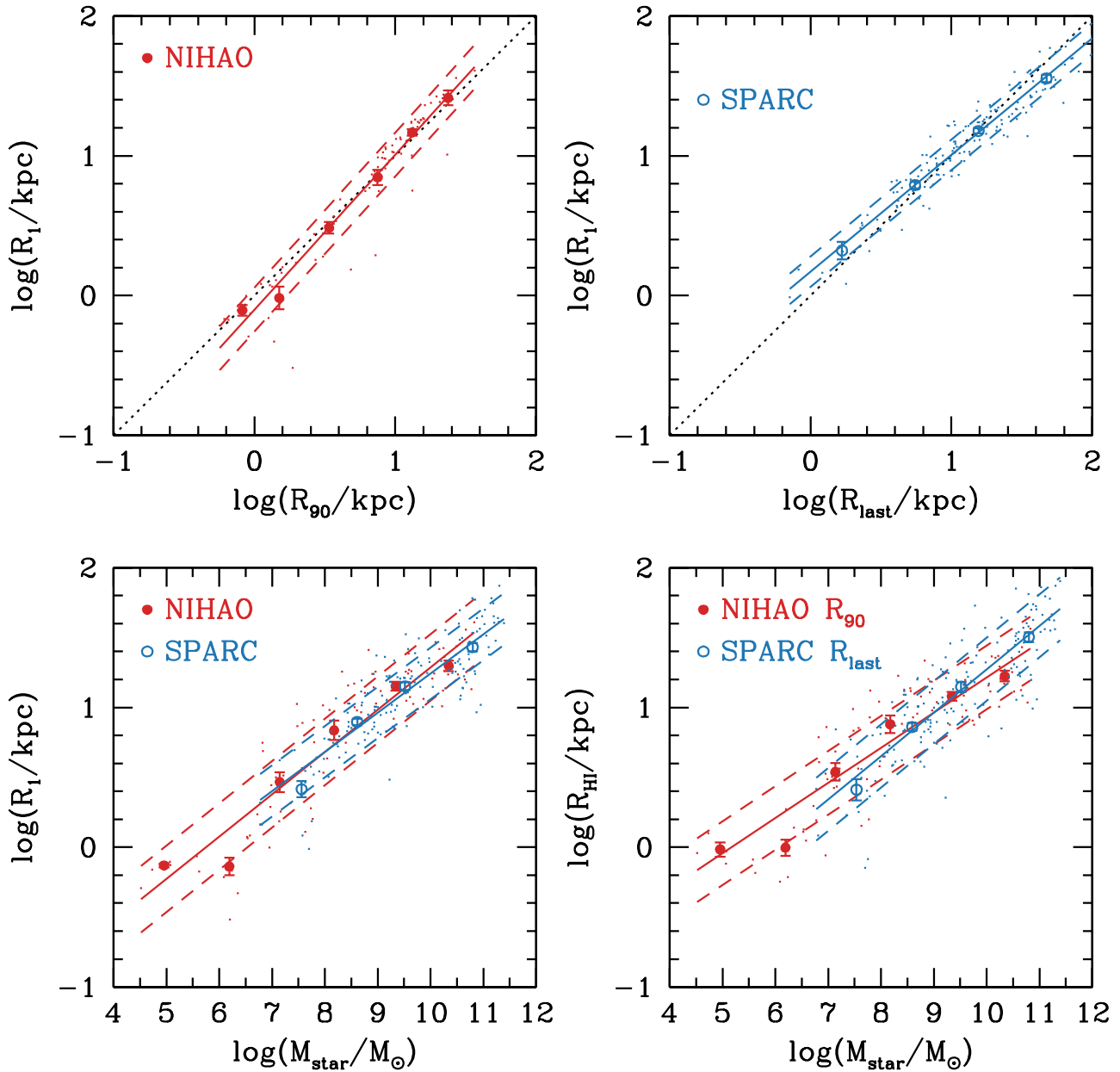}
  \caption{HI size vs stellar mass relation. In the NIHAO simulations
    (red points) sizes  enclose 90\% of the HI mass from a face-on
    projection. In the SPARC observations (blue points) sizes  are the
    last point on the HI rotation curve. Points show individual
    galaxies, circles show mean sizes in bins of HI mass, the error
    bar shows the error on the mean, solid and dashed lines show a
    linear fit and its standard deviation.}
\label{fig:hisizes}
\end{figure*}

\subsection{Skewness of the linewidth distribution}
For observed galaxies we only get a single projection per
galaxy. However, for a sample of galaxies, we can expect a random
distribution of projections, provided we select galaxies by a
parameter that is independent of the projection angle, such as the HI
mass.

Fig.~\ref{fig:wmass} shows the distributions of HI linewidth vs HI
mass.  Single dish observations are from the ALFALFA survey
\citep{Haynes18} and HIPASS \citep{Koribalski04}.  The lower right
panel shows the dependence of $\gamma_1$ measured in bins of HI mass.
The two observational datasets and simulations find consistent
results, for the skewness of both $W_{50}$ (solid lines) and $W_{20}$
(dashed lines).  Namely high mass galaxies are strongly negatively
skewed, close to the value expected for thin rotating disks.  Lower
mass galaxies have higher $\gamma_1$.  This is an observational
signature that lower mass galaxies are more kinematically disordered.
For the lowest mass galaxies the distribution of linewidths is
positively skewed, which is a signature of a floor in the linewidth
from thermal broadening of the HI line.

The distribution of linewidths in bins of HI mass can also be used to
estimate the velocity dispersion of the HI gas, $\sigma_{\rm
  HI}$. This is because for a galaxy disk that is face-on, there will
be zero rotational component projected into the line-of-sight, and
thus the observed linewidth will be a reflection of the disordered
motions of the gas. Assuming the line profiles are Gaussian
$\sigma_{\rm HI}=W_{50}/2.355$.  The minimum linewidth is subject to
observational measurement errors, and may not be representative of the
population, so we take the average of the lowest 5\% of linewidths.
This will give an upper limit to $\sigma_{\rm HI}$ because for a thin
rotating disk the lowest 5\% of inclinations result in a line-of-sight
rotation that is $\sim23\%$ of the edge-on value. 

The results of this exercise are shown with the solid points in
Fig.~\ref{fig:sigmaHI}. Overall there is good agreement between the
observations and simulations.  In detail there are some differences.
Comparing the observations, in the low mass bins HIPASS results in
$\sim 3\kms$ higher $\sigma_{\rm HI}$ than ALFALFA, likely due to the
lower spectral resolution of HIPASS (indicated with horizontal
lines). Except for the highest mass bin, the NIHAO simulations have
slightly higher $\sigma_{\rm HI}$ than ALFALFA. For NIHAO  the red
open squares show the thermal velocity dispersion.  The difference
between the solid red and open red points is thus the contribution of
turbulence (and projected rotation).  Since projection affects both
simulations and observations the slightly higher dispersions in NIHAO
are thus caused by turbulence, likely as a result of the feedback
being too strong.  

The black and grey points show measurements from individual galaxies
using spatially resolved \hi observtions from THINGS \citep{Ianja12}
and LITTLE THINGS \citep{Iorio17}. The solid lines show linear fits
with $3\sigma$ clipping. The slopes are similar, but there is a small
offset of $\simeq 2 \kms$, likely reflecting a systematic in the
measurement techniques. In the mass range $10^7 < M_{\rm HI}  < 10^9
\Msun$ the resolved measurements are in good agreement with the
spatially unresolved linewidth based measurements. Thus in the future
our technique can be used to infer the velocity dispersions of large
samples of spatially unresolved observations.  At higher masses
($M_{\rm HI}> 10^9 \Msun$) the resolved measurements give lower
dispersions than the unresolved measurements. As mentioned above,
projected rotation likely contributes to this difference. In addition,
the linewidth based dispersion includes deviations from a perfectly
axisymmetric disk. For example, a warp will prevent the projected
rotation from being zero at any viewing angle, increasing the implied
dispersion.  Another possible contributing effect is the role of
sample selection. The THINGS survey is not a complete sample of
galaxies, it is possibly biased towards that more rotation dominated
galaxies at these masses than the galaxy population.

In summary, the distribution of linewidths in bins of \hi mass shows
that in both simulations and observations lower mass galaxies are less
rotationally supported than more massive galaxies.  This is driven by
a reduction in the rotation velocity in low mass galaxies rather than
an increase in the gas dispersion, which also decreases in lower mass
galaxies.

\begin{figure*} 
  \includegraphics[width=0.96\textwidth]{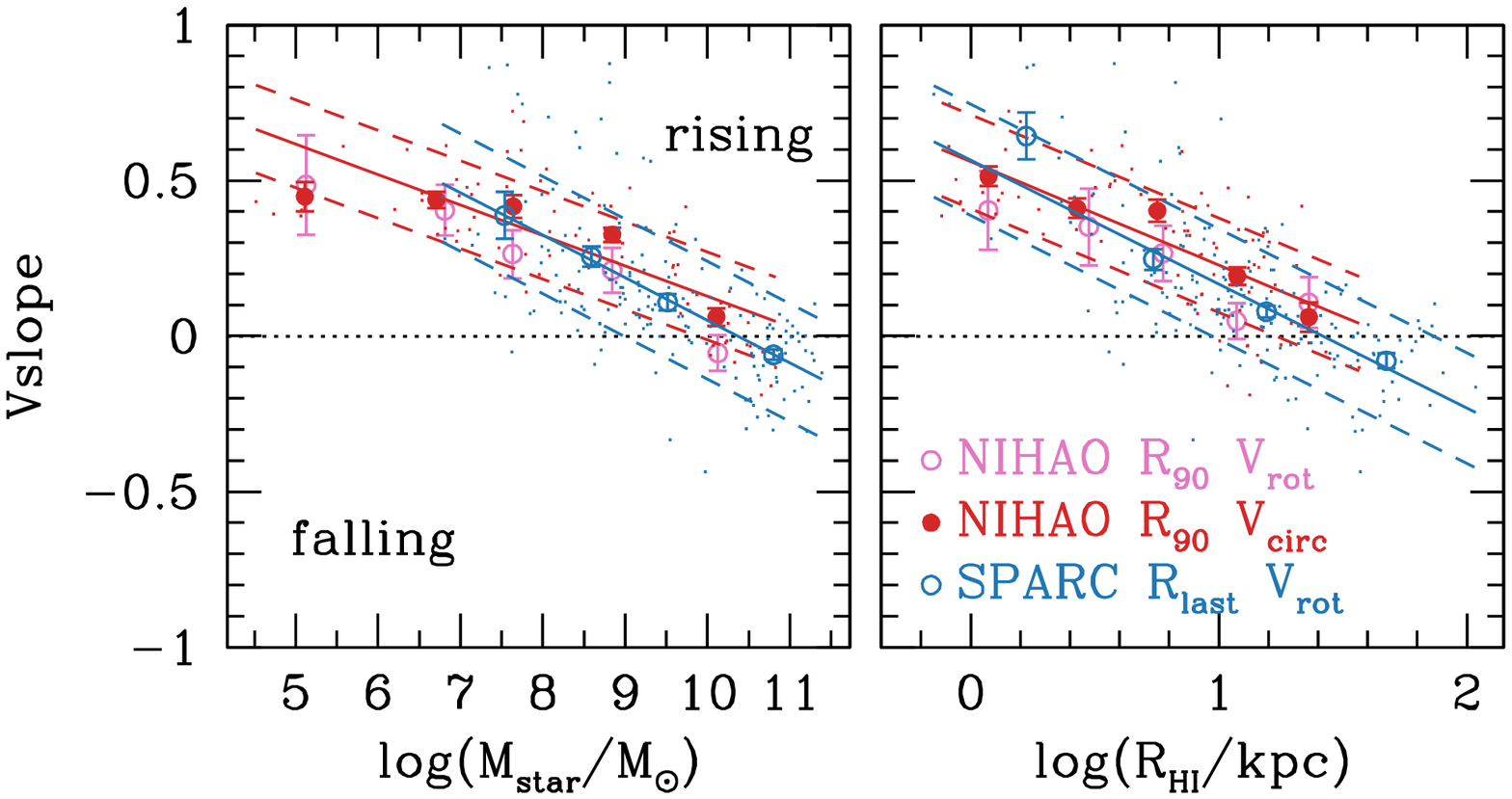}
  \caption{Logarithmic slope of the outer velocity curve vs stellar
    mass (left) and HI radius (right).  In the NIHAO simulations the
    slope is measured between 0.5 and 1.0 $R_{\rm HI}$ on the circular
    velocity curve (red) and the rotation velocity curve (pink).  For
    the observations from SPARC (blue) the slope is measured between
    0.5 and 1.0 $R_{\rm last}$ on the rotation curve.  This shows that
    in both simulations and observations lower mass and smaller
    galaxies have more strongly rising velocity curves.}
\label{fig:rcslope}
\end{figure*}

\subsection{HI sizes}
We showed previously in \citet{Maccio16} that the NIHAO galaxies match
two HI size vs HI mass relations (the exponential scale length, and
the radius where the surface density reaches $1 \Msun {\rm
  pc}^{-2}$). We are interested in the relation between HI size and
halo mass, but halo masses are not directly observable.  In the NIHAO
simulations, stellar mass is the galaxy observable that is most
strongly correlated with halo mass (correlation coefficient of
0.98). The scatter in the stellar mass vs halo mass relation is small
($\sim 0.2$ dex) both in theory \citep{Dutton17,Matthee17}, and
observations \citep{More11,Reddick13, Moster18}. Here we use galaxies
from the SPARC survey \citep{Lelli16} as this is the largest
compilation of resolved HI rotation curves with $3.6 \mu m$ Spitzer
imaging (which gives the most reliable stellar mass tracer).  

Fig.~\ref{fig:hisizes} shows HI size vs stellar mass relations and a
comparison between various HI size definitions.  The upper panels show
that $R_1$ (the radius where the HI surface density drops to $1 \Msun
\pc$) is closely related to $R_{90}$ in the NIHAO simulations and
$R_{\rm last}$ in the SPARC observations.  The lower left panel shows
a very good agreement between the $R_1$ vs stellar mass relations in
NIHAO (red) and SPARC (blue).  Fits to the size-mass relations are
given in Table~\ref{tab:fits}.

While the observations are not a complete sample of nearby galaxies
the comparison is a good place to start.  It contradicts the claim by
\citep{Trujillo-Gomez18} that the HI distribution in NIHAO galaxies is
too compact compared to observations. This was suggested by
\citep{Trujillo-Gomez18} as a reason for the low
$W_{50}/2\VmaxDMO$ ratio in NIHAO dwarf galaxies. As shown in
(Fig.~\ref{fig:vsvmax}), the HI in NIHAO dwarf galaxies is less
extended relative to the dark matter halo than in more massive
galaxies. So this does contribute to the low $W_{50}/2\VmaxDMO$ ratios
in dwarfs. However, rather than being an artifact of 
NIHAO, this appears to be a feature of real galaxies.

\subsection{Outer HI rotation curve slopes}

Fig.~\ref{fig:rcslope} shows the logarithmic slope ($\Delta \log V /
\Delta \log R$) of the outer rotation curve vs the stellar mass (left)
and the HI radius (right). Points show individual galaxies, while
symbols with error bars show the mean and error on the mean in bins of
mass or radius. For the observations the slope is measured between 0.5
and 1.0 $R_{\rm last}$  on the rotation curve, for the simulations the
fiducial slope is measured between 0.5 and 1.0 $R_{\rm HI}$ on the
circular velocity curve (red filled circles), and on the rotation
curve (pink open circles).

The simulations are in good agreement with the observations, except
that the observed relations have slightly larger scatter, plausibly
due to larger measurement errors.  We see that high mass or large
galaxies have, on average, flat outer rotation curves at the HI
radius. As we go to lower stellar masses and smaller sizes the outer
rotation curves are progressively rising.  Notice that even though
both observations and simulations include cored dark matter density
profiles in dwarf galaxies, at large radii the typical outer
rotation curve slope is $\sim 0.5$ corresponding to a $\rho\propto
r^{-1}$ density profile.  When the rotation curve is rising, the
half-linewidth $W_{50}/2$ will neccesarily underestimate the maximum
rotation velocity.

\subsection{Axis ratios}
\begin{figure} 
  \includegraphics[width=0.48\textwidth]{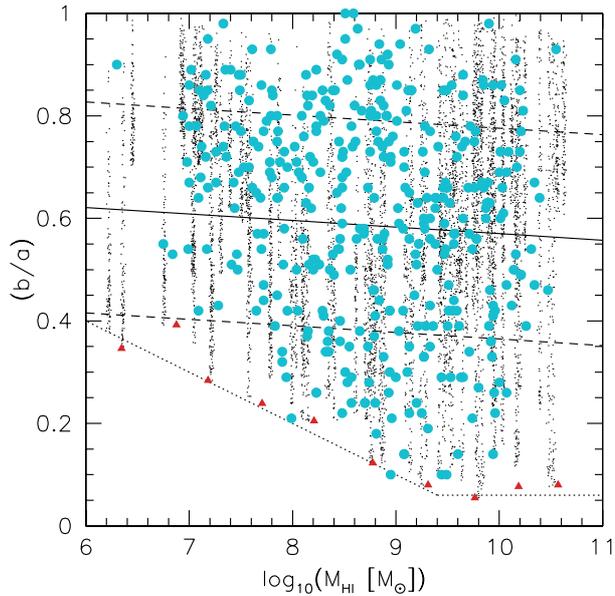}
  \caption{Dependence of projected HI axis ratio on HI mass. For NIHAO
    simulations each galaxy has 100 projections shown (black points).
    The red triangles show the minimum axis ratio in bins of mass
    showing a deficit of small axis ratios in low mass galaxies.  This
    is consistent with observations (cyan circles) from
    \citet{Wang16}.}
\label{fig:axis}
\end{figure}

Fig.~\ref{fig:axis} shows the relations between projected axis ratio
and HI mass. Each galaxy has 100 projections shown. For clarity we
have randomly shifted the HI masses by a small amount for each
projection. There is a weak trend, similar to that between minimum
axis ratio and halo velocity shown in Fig.~\ref{fig:vsvmax}, with a
scatter of 0.20.  The average projected axis ratio is about 0.6,
compared to 0.5 for an idealized thin disk.  The clearest testable
prediction is the minimum axis ratio is larger in lower mass
galaxies. The most massive galaxies can have disks as thin as
0.05. Below a mass of $M_{\rm HI} \sim 10^9\Msun$ the minimum axis
ratio steadily increases reaching 0.4 at $M_{\rm HI} \sim 10^6\Msun$
The lower envelope (dotted line in Fig.~\ref{fig:axis}) is given by
\begin{equation}
  (b/a)_{\rm min} = 0.06  + 0.106( \log_{10}(M_{\rm HI}) -9.4)  
\end{equation}
for $\log_{10}(M_{\rm HI}) \le 9.4$, and   $(b/a)_{\rm min} = 0.06$ otherwise.

The cyan circles show observed galaxies from the nearby Universe from
\citet{Wang16}.  This study combines \hi data from several different
projects so is representative of a wide variety of galaxy types and
environments: Atlas3D \citep{Serra12,Serra14}, Bluedisk
\citep{Wang13}, Faint Irregular Galaxies GMRT Survey (FIGGS)
\citep{Begum08}.  LITTLE THINGS \citep{Hunter12}, Local Volume HI
Survey (LVHIS) \citep{Koribalski18},  starbursting dwarf galaxies
\citep{Lelli14}, The HI Nearby Galaxy Survey (THINGS)
\citep{Walter08}, Ursa Major cluster \citep {Verheijen01}, Void Galaxy
Survey (VGS) \citep{Kreckel12}, VLA Imaging of Virgo Spirals in Atomic
Gas (VIVA) \citep{Chung09}, Westerbork HI survey of spiral and
irregular galaxies (WHISP) \citep{Swaters02, Noordermeer05}.

For the vast majority of galaxies (88\%) the axis ratio  is determined
by \citet{Wang16} from the \hi maps, based on the second order moments
of the pixel distributions where $\Sigma_{\rm HI} > 1 \Msun \pc$. For
the galaxies in VIVA (36) and Atlas3D (8) axis ratios are from  tilted
ring fits to the velocity fields.  Galaxies that are poorly resolved
(HI radius less than the beam major axis) have been excluded, though
this does not guarantee the axis ratios are well resolved in all
galaxies.

The observations and simulations have similar distributions of axis
ratios, in particular there are no observed galaxies in the lower left
corner where our simulations predict there to be none.  This appears
to confirm the lack of dwarf galaxies with thin HI disks.  Future
observations of a complete sample of galaxies with higher spatial
resolution are needed to conclusively confirm the lack of thin dwarf
galaxies.

\section{Summary}
\label{sec:sum}

We use a sample of 85 galaxies simulated in a LCDM cosmology from the
NIHAO project to investigate why HI linewidths systematically
underpredict the maximum dark halo circular velocities in dwarf
galaxies (Fig.~\ref{fig:wvmax}).  We trace this to two primary effects.

\begin{itemize}
\item Lower mass galaxies are less rotationally supported. This is
  confirmed observationally from the skewness of linewidths in bins of HI mass in both
  ALFALFA and HIPASS observations (Fig.~\ref{fig:wmass}).

\item The HI distributions are less extended (relative to the dark
  matter halo) in dwarf galaxies, so that the rotation curves are
  still rising at the last measured data point, in
  agreement with observations (Fig.~\ref{fig:rcslope}). 
\end{itemize}

The HI profile shape parameterized by $W_{50}/W_{20}$ decreases in
lower mass galaxies (Fig.~\ref{fig:ww}) consistent with both these two
effects. An additional observational test is in the distribution of HI
axis ratios. In particular, in the NIHAO simulations the minimum axis
ratio is larger in lower mass galaxies (Fig.~\ref{fig:axis}).

In our simulations the HI kinematics are an inhomogeneous
population. There is a significant range of rotational support and HI
extent at any given halo or galaxy mass. This variation drives the
variation in $W_{50}/2\VmaxDMO$ (Fig.~\ref{fig:wvmax2}).  Thin and
extended rotating HI disks exist in our simulations at all halo
masses, but they are not a fair sample.  This implies that one cannot
use a sample of well ordered rotating disks to interpret the HI
linewidths of large unbiased samples of galaxies, as is sometimes done
\citep[e.g.,][]{Papastergis17, Trujillo-Gomez18}.

The implied linewidth velocity function from the NIHAO simulations has
a shallow slope ($\simeq -1$) at low velocities ($10 < W_{50}/2 <
80\,\kms$), consistent with observations (Fig.~\ref{fig:vf4}). Thus
we conclude that the apparent discrepancy between the
predictions of the LCDM cosmological model and observations as
highlighted by previous authors \citep{Papastergis11,Klypin15} is due
to their incorrect assumption that $W_{50}/2 \simeq \VmaxDMO$. 

We look forward to the next generation of blind HI line surveys, APERTIF
\citep{Verheijen08} and ASKAP \citep{Johnston08} that will provide
higher spatial resolution data with which to further
test our simulations and thus the LCDM model. 

\section*{Acknowledgments}
We thank the anonymous referee for a detailed and constructive report.
This research was carried out on the High Performance Computing
resources at New York University Abu Dhabi; on the  {\sc theo} cluster
of the Max-Planck-Institut f\"ur Astronomie and on the {\sc hydra}
clusters at the Rechenzentrum in Garching.
The authors gratefully acknowledge the Gauss Centre for Supercomputing
e.V. (www.gauss-centre.eu) for funding this project by providing
computing time on the GCS Supercomputer SuperMUC at Leibniz
Supercomputing Centre (www.lrz.de).
AO is funded by the Deutsche Forschungsgemeinschaft (DFG, German  
Research Foundation) -- MO 2979/1-1.


\end{document}